\documentclass[conference]{IEEEtran}
\usepackage[dvips]{graphicx}
\usepackage[cmex10]{amsmath}
\usepackage{multirow}
\usepackage{epsfig}
\usepackage{mathrsfs}
\usepackage{amssymb}
\usepackage{comment}
\usepackage{bm}
\usepackage{url}
\usepackage{array}
\usepackage{amsthm}
\usepackage{color}
\usepackage{blkarray}
\usepackage{fancyhdr}
\usepackage{enumerate}
\usepackage[lined,boxed,commentsnumbered,ruled,linesnumbered]{algorithm2e}

\renewcommand{\thispagestyle}[1]{} 

\def\clr{\color{black}}

\theoremstyle{definition}
\newtheorem{theorem}{Theorem}[section]
\newtheorem{lemma}[theorem]{Lemma}
\newtheorem{corollary}[theorem]{Corollary}
\newtheorem{proposition}[theorem]{Proposition}

\newcommand{\OmegaCSP}{\Omega^{\mbox{\tiny CSP}}}
\newcommand{\OmegaUP}{\Omega^{\mbox{\tiny UP}}}

\newcommand{\MSC}{\mbox{MSC}}
\newcommand{\GSC}{\mbox{GSC}}

\newcommand{\SCSP}{S^*_{\mbox{\tiny CSP}}}
\newcommand{\SUP}{S^*_{\mbox{\tiny UP}}}

\newcommand{\SCSPi}{S_{\mbox{\tiny CSP}}^{\mbox{\tiny inner}}}
\newcommand{\SCSPo}{S_{\mbox{\tiny CSP}}^{\mbox{\tiny outer}}}
\newcommand{\SUPi}{S_{\mbox{\tiny UP}}^{\mbox{\tiny inner}}}
\newcommand{\SUPo}{S_{\mbox{\tiny UP}}^{\mbox{\tiny outer}}}

\DeclareMathAlphabet{\mathcal}{OMS}{cmsy}{m}{n}

\ifCLASSINFOpdf

\else

\fi

\usepackage[tight,footnotesize]{subfigure}

\begin{document}
\pagestyle{fancy}
\IEEEoverridecommandlockouts

\lhead{\textit{Technical Report, IBM T. J. Watson Research Center, Yorktown, NY, USA, March, 2015.}}
\rhead{} 
%
\title{Failure Localization Capability: Theorem Proof and Evaluation}
\author{\IEEEauthorblockN{Liang Ma\IEEEauthorrefmark{2}, Ting He\IEEEauthorrefmark{2}, Ananthram Swami\IEEEauthorrefmark{4}, Don Towsley\IEEEauthorrefmark{1}, and Kin K. Leung\IEEEauthorrefmark{3}\\}
\IEEEauthorblockA{\IEEEauthorrefmark{2}IBM T. J. Watson Research Center, Yorktown, NY, USA. Email: \{maliang, the\}@us.ibm.com\\
\IEEEauthorrefmark{4}Army Research Laboratory, Adelphi, MD, USA. Email: ananthram.swami.civ@mail.mil\\
\IEEEauthorrefmark{1}University of Massachusetts, Amherst, MA, USA. Email: towsley@cs.umass.edu\\
\IEEEauthorrefmark{3}Imperial College, London, UK. Email: kin.leung@imperial.ac.uk
}\vspace{2em}
}

\maketitle

\IEEEpeerreviewmaketitle

\section{Introduction}
Selected theorem proofs in \cite{Ma15TON} and additional simulation results are presented in detail in this report. We first list the theorems in Section~\ref{sect:theorems}, and then give the proofs in Section~\ref{sect:proofs}. In the end, we show some additional simulation results for extensive comparison of network failure localization capabilities under various probing mechanisms in Section~\ref{sect:evaluation}. See the original paper \cite{Ma15TON} for terms and definitions.

\section{Theorems}
\label{sect:theorems}

\begin{lemma}\label{lem:abstract condition}
Set $S$ is $k$-identifiable:
\begin{enumerate}
\item[a)] if for any failure set $F$ with $|F|\leq k$ and any node $v$ with $v\in S\setminus F$, $\exists$ $p\in P$ traversing $v$ but none of the nodes in $F$;
\item[b)] only if for any failure set $F$ with $|F|\leq k-1$ and any node $v$ with $v\in S\setminus F$, $\exists$ $p\in P$ traversing $v$ but none of the nodes in $F$.
\end{enumerate}
\end{lemma}

\begin{corollary}
\label{coro:maximum_set_lower_bound}
Let $S''(k):=\{v\in N:\exists$ path in $P$ traversing $v$ but none of the nodes in each failure set $F$ with $v\notin F$ and $|F|\leq k\}$. Then $S''(k)\subseteq S^*(k)$.
\end{corollary}

\begin{theorem}[$k$-identifiability under CSP]\label{thm:k-identifiability, CSP}
Set $S$ is $k$-identifiable under CSP:
\begin{enumerate}
\item[a)] if $\Gamma_{\mathcal{G}^*}(S,m')\geq k+2$, and $\min_{m\in M}\Gamma_{\mathcal{G}_m}(S,m')\geq k+1$ ($k\leq \sigma-2$);
\item[b)] only if $\Gamma_{\mathcal{G}^*}(S,m')\geq k+1$, and $\min_{m\in M}\Gamma_{\mathcal{G}_m}(S,m')\geq k$ ($k\leq \sigma-1$).
\end{enumerate}
\end{theorem}

\begin{lemma}\label{lem:connectivity of G*}
Each connected component in $\mathcal{G}-V'$ that contains a node in $S$ has a monitor for any set $V'$ of up to $q$ ($q\leq \sigma-1$) non-monitors if and only if $\Gamma_{\mathcal{G}^*}(S, m')\geq q+1$.
\end{lemma}

\begin{lemma}\label{lem:k-identifiability, CSP}
Set $S$ is $k$-identifiable under CSP:
\begin{enumerate}
\item[a)] if for any node set $V'$, $|V'|\leq k+1$, containing at most one monitor, each connected component in $\mathcal{G}-V'$ that contains a node in $S$ has a monitor;
\item[b)] only if for any node set $V'$, $|V'|\leq k$, containing at most one monitor, each connected component in $\mathcal{G}-V'$ that contains a node in $S$ has a monitor.
\end{enumerate}
\end{lemma}

\begin{lemma}\label{lem:connectivity of G_i}
The following two conditions are equivalent:
\begin{enumerate}
  \item[(1)] Each connected component in $\mathcal{G}-V'$ that contains a node in $S$ has a monitor for any set $V'$ consisting of monitor $m$ ($m\in M$) and up to $q$ ($q\leq \sigma-1$) non-monitors;
  \item[(2)] $\Gamma_{\mathcal{G}_m}(S,m')\geq q+1$.
\end{enumerate}
\end{lemma}

\begin{proposition}\label{prop:sigma-identifiability, CSP}
Set $S$ is $\sigma$-identifiable under CSP if and only if each non-monitor in $S$ has at least two monitors as neighbors.
\end{proposition}

\begin{proposition}\label{prop:(sigma-1)-identifiability, CSP}
Set $S$ is $(\sigma-1)$-identifiable under CSP if and only if (i) all nodes in $S$ have at least two monitors, or (ii) all nodes in $N$ but $v$ ($v\in S$) have at least two monitor neighbors and $v$ has one monitor and all nodes in $N\setminus \{v\}$ as neighbors.
\end{proposition}

\begin{theorem}[$k$-identifiability under UP]\label{thm:k-identifiability, UP}
Set $S$ is $k$-identifiable under UP with measurement paths $P$: \begin{enumerate}
\item[a)] if $\MSC(v)\geq k+1$ for any node $v$ in $S$ ($k\leq \sigma-1$);
\item[b)] only if $\MSC(v)\geq k$ for any node $v$ in $S$ ($k\leq \sigma$).
\end{enumerate}
\end{theorem}

\begin{proposition}
\label{prop:sigma-identifiability_UP}
Set $S$ is $\sigma$-identifiable under UP if and only if $MSC(v)=\sigma$ for any node $v$ in $S$, i.e., each node in $S$ is on a $2$-hop path.
\end{proposition}

\begin{corollary}
\label{coro:Maximum_k_identifiable_set-CSP}
Let $\SCSPo(k):=\{v \in N:\pi_v\geq k\}$, and $\SCSPi(k):=\{v \in N:\pi_v\geq k+1\}$. The maximum $k$-identifiable set under CSP ($k\leq \sigma-1$), denoted by $\SCSP(k)$, is bounded by $\SCSPi(k)\subseteq\SCSP(k)\subseteq\SCSPo(k)$.
\end{corollary}

\begin{corollary}
\label{coro:Maximum_sigma-1_identifiable_set-CSP}
$\SCSP(\sigma-1)=\{v\in N:v$ has at least two monitor neighbors$\}\cup \widetilde{S}$. Set $\widetilde{S}$ contains one and only one non-monitor $w$ if all nodes in $N$ but $w$ have at least two monitor neighbors and $w$ has one monitor and all nodes in $N\setminus \{w\}$ as neighbors; otherwise, $\widetilde{S}=\emptyset$.
\end{corollary}

\begin{corollary}
\label{coro:Maximum_k_identifiable_set-UP}
Let $\SUPo(k):=\{v \in N:\MSC(v)\geq k\}$ and $\SUPi(k):=\{v \in N:\MSC(v)\geq k+1\}$ with measurement paths $P$. The maximum $k$-identifiable set under UP ($k\leq \sigma-1$), denoted by $\SUP(k)$, is bounded by $\SUPi(k)\subseteq\SUP(k)\subseteq\SUPo(k)$.
\end{corollary}

\begin{theorem}[Maximum Per-node Identifiability under CSP]\label{thm:Omega_CSP}
If $\pi_v \leq \sigma-2$, the maximum identifiability of a non-monitor $v$ under CSP is bounded by $\pi_v - 1 \leq \OmegaCSP(v) \leq \pi_v$.
\end{theorem}

\begin{theorem}[Maximum Per-node Identifiability under UP]\label{thm:Omega_UP}
The maximum identifiability of a non-monitor $v$ under UP with measurement paths $P$ is bounded by $\MSC(v) - 1 \leq \OmegaUP(v) \leq \MSC(v)$.
\end{theorem}

\section{Proofs}
\label{sect:proofs}

\subsection{Proof of Lemma~\ref{lem:connectivity of G*}}
If the first condition holds, then each connected component in $\mathcal{G}-M-V'$ that contains a node in $S$ has a neighbor of a monitor. Since these neighbors are connected to $m'$ in $\mathcal{G}^*-V'$, each node $v$ with $v\in S$ in $\mathcal{G}^*-V'$ is connected to $m'$. If the first condition is violated, i.e., there exists a connected component in $\mathcal{G}-M-V'$ that contains a node in $S$ has no neighbor of any monitor, then this component must be disconnected from $m'$ in $\mathcal{G}^*-V'$.
$\hfill\blacksquare$

\subsection{Proof of Lemma~\ref{lem:k-identifiability, CSP}}
Suppose condition (a) holds, and consider a candidate failure set ${F}$, $|{F}|\leq k$ and a non-monitor $v\in S\setminus {F}$. We argue that $v$ must have two simple \emph{vertex disjoint} paths to monitors in $\mathcal{G}-{F}$, and thus concatenating these paths provides a monitor-monitor simple path that traverses $v$ but not ${F}$, satisfying the abstract sufficient condition in Lemma~\ref{lem:abstract condition}. Indeed, if such paths do not exist, i.e., $\exists$ a (monitor or non-monitor) node $w$ ($w\neq v$) that resides on all paths from $v$ to monitors in $\mathcal{G}-{F}$, then $v$ will be disconnected from all monitors in $\mathcal{G}-{F}-\{w\}$, i.e., the connected component containing $v$ in $\mathcal{G}-V'$, where $V'={F}\cup \{w\}$, has no monitor, contradicting condition (a).

Suppose condition (b) does not hold, i.e., there exists a non-monitor $v$ in $S$, a (monitor or non-monitor) node $w$, and a set of up to $k-1$ non-monitors ${F}$ ($v\neq w$ and $v,w\not\in{F}$) such that the connected component containing $v$ in $\mathcal{G}-V'$, $V'={F}\cup \{w\}$, contains no monitor. Then any path (if any) from $v$ to monitors in $\mathcal{G}-{F}$  must traverse $w$, which means no monitor-monitor simple path in $\mathcal{G}-{F}$ will traverse $v$ (as any monitor-monitor path traversing $v$ must form a cycle at $w$). Therefore, if node $v$ fails, the failure cannot be identified in $\mathcal{G}-{F}$.
$\hfill\blacksquare$

\subsection{Proof of Lemma~\ref{lem:connectivity of G_i}}
The proof is similar to that of Lemma~\ref{lem:connectivity of G*}. If the first condition holds, then each connected component in $\mathcal{G}-M-F$ for $F= V'\setminus \{m\}$ contains a node in $\mathcal{N}(M\setminus \{m\})$. Thus each node in $S\setminus V'$ is connected to $m'$ in $\mathcal{G}_m - F$. If the first condition is violated, then there exists a connected component in $\mathcal{G}-M-F$ that contains a node in $S$ does not contain any node in $\mathcal{N}(M\setminus \{m\})$, and thus this component containing nodes in $S\setminus V'$ must be disconnected from $m'$ in $\mathcal{G}_m - F$. Hence, the first condition is equivalent to the second condition.
$\hfill\blacksquare$

\subsection{Proof of Proposition~\ref{prop:sigma-identifiability, CSP}}
If each node in $S$ has at least two monitors as neighbors, then their states can be determined independently by cycle-free 2-hop probing between monitors, and thus $S$ is $\sigma$-identifiable. On the other hand, suppose $\exists$ a non-monitor $v$ in $S$ with zero or only one monitor neighbor. Then $\nexists$ simple paths going through $v$ without traversing another non-monitor, and hence the state of $v$ cannot be determined if all the other non-monitors fail.
$\hfill\blacksquare$

\subsection{Proof of Proposition~\ref{prop:(sigma-1)-identifiability, CSP}}
\emph{Necessity:} Suppose that $S$ is $(\sigma-1)$-identifiable under CSP. If it is also $\sigma$-identifiable, then each node in $S$ must have at least two monitor neighbors according to {\clr Proposition}~\ref{prop:sigma-identifiability, CSP}. Otherwise, we have $\Omega(S)=\sigma-1$. In this case, $\exists$ at least one node in $S$, denoted by $v$, with at most one monitor neighbor. Let $\mathcal{N}(v)$ denote all neighbors of $v$ including monitors. Suppose that $v$ has $\lambda$ neighbors (i.e., $|\mathcal{N}(v)|=\lambda$). Then there are two cases: (i) $\mathcal{N}(v)$ contains a monitor, denoted by $\widetilde{m}$; (ii) all nodes in $\mathcal{N}(v)$ are non-monitors. In case (i), the sets $F_1=\mathcal{N}(v)\setminus \{\widetilde{m}\}$ and $F_2=F_1\cup \{v\}$ are not distinguishable because $\nexists$ monitor-to-monitor simple paths traversing $v$ but not nodes in $F_1$. In case (ii), the sets $F_1=\mathcal{N}(v)\setminus \{w\}$ (where $w$ is an arbitrary node in $\mathcal{N}(v)$) and $F_2=F_1\cup \{v\}$ are not distinguishable as all monitor-to-monitor simple paths traversing $v$ must go through at least one node in $F_1$. Based on (i--ii), we conclude that $\Omega(\mathcal{G})\leq \lambda-1$, where $\lambda$ is the degree of any node in $S$ with at most one monitor neighbor. For $\Omega(\mathcal{G})=\sigma-1$, we must have $\lambda\geq \sigma$, which can only be satisfied if all such nodes in $S$ have one monitor and all the other non-monitors in $N$ as neighbors. Moreover, if there are two such nodes $v$ and $u$ in $S$, then the sets $F\cup \{v\}$ and $F\cup \{u\}$, where $F=N\setminus \{v,\: u\}$, are not distinguishable as all monitor-to-monitor simple paths traversing $v$ must go through $F$ or $u$ and vice versa. Therefore, such node, $v$, in $S$ must be unique. Now suppose $\exists$ node $z$ ($z\in N\setminus S$) which has no or only one monitor neighbor. Then failure sets $F\cup \{v\}$ and $F\cup \{z\}$, where $F=N\setminus \{v,\: z\}$, are not distinguishable as $P_v=P_z$ in $\mathcal{G}-F$. Thus, all nodes in $N\setminus v$ must have two monitor neighbors.

\emph{Sufficiency:} If each node in $S$ has at least two monitor neighbors, then $S$ is $\sigma$-identifiable (hence also $(\sigma-1)$-identifiable) according to {\clr Proposition}~\ref{prop:sigma-identifiability, CSP}. If condition in Proposition~\ref{prop:(sigma-1)-identifiability, CSP} holds and node $v$ is the only node in $S$ which has less than two monitor neighbors, then for any two failure sets $F_1$ and $F_2$ with $|F_i|\leq \sigma-1$ ($i=1,\: 2$) and $F_1\cap S \neq F_2\cap S$, there are two cases: (i) $F_1$ and $F_2$ differ on a non-monitor other than $v$; (ii) $F_1$ and $F_2$ only differ on $v$. In case (i), since the states of all non-monitors other than $v$ can be independently determined, $F_1$ and $F_2$ are distinguishable. In case (ii), suppose that $F_1 = F\cup \{v\}$ and $F_2 = F$ for $F\subseteq N\setminus \{v\}$. Since $|F_1|\leq \sigma-1$, $|F|\leq \sigma-2$ and $\exists$ a non-monitor $w\in (N\setminus \{v\})\setminus F$. We know that $v$ is a neighbor of $w$ (as $v$ is a neighbor of all the other non-monitors) and $w$ is a neighbor of a monitor $m$ other than $\widetilde{m}$ (as it has at least two monitor neighbors). Thus, $\widetilde{m}vwm$ is a monitor-to-monitor simple path traversing $v$ but not $F$, whose measurement can distinguish $F_1$ and $F_2$. Therefore, $S$ is $(\sigma-1)$-identifiable under CSP.
$\hfill\blacksquare$

\subsection{Proof of Theorem~\ref{thm:k-identifiability, UP}}
Suppose condition (a) holds. Then for any candidate failure set $F$ with $|F|\leq k$ and any node $v$ with $v\in S\setminus F$, there must be a path in $P_v$ that is not in $\bigcup_{w\in F}P_w$, i.e., traversing $v$ but not $F$, which satisfies the abstract sufficient condition in Lemma~\ref{lem:abstract condition}.

Suppose condition (b) does not hold, i.e., there exists node $v$ in $S$ and a set of non-monitors $V'$ with $|V'|\leq k-1$ and $v\not\in V'$, such that $P_v \subseteq \bigcup_{w\in V'}P_w$. Then given failures of all nodes in $V'$, the state of $v$ has no impact on observed path states and is thus unidentifiable, violating the abstract necessary condition in Lemma~\ref{lem:abstract condition}.
$\hfill\blacksquare$

\subsection{Proof of Proposition~\ref{prop:sigma-identifiability_UP}}
Similar to the proof of Proposition~\ref{prop:sigma-identifiability, CSP}, if each node in $S$ is on a $2$-hop path, then their states can be determined independently, and thus $S$ is $\sigma$-identifiable under UP. On the other hand, suppose $\exists$ a non-monitor $v$ in $S$ which is not on any $2$-hop paths. Then the state of $v$ cannot be determined if all the other non-monitors fail.
$\hfill\blacksquare$

\subsection{Proof of Corollary~\ref{coro:Maximum_k_identifiable_set-CSP}}
For each node $q$ in $\SCSPi(k)$, $\exists$ a path traversing $q$ but none of the nodes in each failure set $F$ with $q\notin F$ and $|F|\leq k$ ($k\leq \sigma-1$). Thus, by Corollary~\ref{coro:maximum_set_lower_bound}, $\SCSPi(k)\subseteq\SCSP(k)$. On the other hand, all nodes in $\SCSPo(k)$ are $(k-1)$-identifiable, and all nodes in $N\setminus \SCSPo(k)$ are at most $(k-1)$-identifiable, i.e., not $k$-identifiable. Thus, all $k$-identifiable nodes are within $\SCSPo(k)$.
$\hfill\blacksquare$

\subsection{Proof of Corollary~\ref{coro:Maximum_sigma-1_identifiable_set-CSP}}
First, all non-monitors with at least two monitor neighbors are $\sigma$-identifiable, thus included in $\SCSP(\sigma-1)$. For node $w$ (if any) in $\widetilde{S}$, it satisfies the necessary and sufficient conditions in Proposition~\ref{prop:(sigma-1)-identifiability, CSP}, and thus $w$ is $(\sigma-1)$-identifiable. Moreover, by Proposition~\ref{prop:(sigma-1)-identifiability, CSP}, no nodes outside $\{v\in N:v$ has at least two monitor neighbors$\}\cup \widetilde{S}$ are $(\sigma-1)$-identifiable. Therefore, $\SCSP(\sigma-1)=\{v\in N:v$ has at least two monitor neighbors$\}\cup \widetilde{S}$.
$\hfill\blacksquare$

\subsection{Proof of Corollary~\ref{coro:Maximum_k_identifiable_set-UP}}
Similar to the proof for Corollary~\ref{coro:Maximum_k_identifiable_set-CSP}, for each node $q$ in $\SUPi(k)$, $\exists$ a path in $P$ traversing $q$ but none of the nodes in each failure set $F$ with $q\notin F$ and $|F|\leq k$ ($k\leq \sigma - 1$). Thus, by Corollary~\ref{coro:maximum_set_lower_bound}, $\SUPi(k)\subseteq\SUP(k)$. On the other hand, all nodes in $N\setminus \SUPo(k)$ are not $k$-identifiable. Thus, all $k$-identifiable nodes are within $\SUPo(k)$.
$\hfill\blacksquare$

\subsection{Proof of Theorem~\ref{thm:Omega_CSP}}
$S=v$ satisfies the condition in Theorem~\ref{thm:k-identifiability, CSP}~(a) for $k = \pi_v-1$. Thus, $\OmegaCSP(v) \geq \pi_v-1$. However, $S=v$ violates the condition in Theorem~\ref{thm:k-identifiability, CSP}~(b) for $k = \pi_v+1$ (which requires $\pi_v+1 \leq \sigma-1$), i.e., $\OmegaCSP(v) \neq \pi_v+1$. Thus, $\OmegaCSP(v) \leq \pi_v$.
$\hfill\blacksquare$

\subsection{Proof of Theorem~\ref{thm:Omega_UP}}
There are two cases for $\MSC(v)$: (i) $\MSC(v)\leq \sigma-1$; (ii) $\MSC(v)= \sigma$. In case (i), $v$ is $(\MSC(v)-1)$-identifiable by Theorem~\ref{thm:k-identifiability, UP}-(a). Meanwhile, $v$ is not $(\MSC(v)+1)$-identifiable by Theorem~\ref{thm:k-identifiability, UP}~(b) (which requires $\MSC(v)+1 \leq \sigma$ when applying Theorem~\ref{thm:k-identifiability, UP}-(b)). Together, they imply the bounds on $\OmegaUP(v)$. For case (ii), node $v$ is on a $2$-hop measurement path, whose state can be determined independently; therefore, $\OmegaUP(v)=\MSC(v)=\sigma$ in this case.
$\hfill\blacksquare$

\section{Extensive Evaluation of Failure Localization Capability}
\label{sect:evaluation}

\subsection{Synthetic Topologies}\label{subsubsec:Random Topologies}

For synthetic topologies, we consider four widely used random graph models: Erd\"{o}s-R\'{e}nyi (ER) graphs, Random Geometric (RG) graphs, Barab\'{a}si-Albert (BA) graphs, and Random Power Law (RPL) graphs. We randomly generate graph realizations of each model\footnote{All realizations are guaranteed to be connected, as we discard disconnected realizations in the generation process.}, with each realization containing $20$ nodes (i.e., $|V|=20$). The generated graphs are then used to evaluate the impact of probing mechanisms.

\emph{\textbf{Erd\"{o}s-R\'{e}nyi (ER) graph}}:
The ER graph \cite{ErdosRenyi60} is generated by independently connecting each pair of nodes by a link with a fixed probability $p$. The result is a purely random topology where all graphs with an equal number of links are equally likely to be selected (note that the number of nodes is a predetermined parameter).

\emph{\textbf{Random Geometric (RG) graph}}:
The RG graph \cite{GuptaKumar99} is frequently used to model the topology of wireless ad hoc networks. It generates a random graph by first randomly distributing nodes in a unit square, and then connecting each pair of nodes by a link if their distance is no larger than a threshold $d_c$, which denotes the node communication range. The resulting topology contains well-connected sub-graphs in densely populated areas and poorly-connected sub-graphs in sparsely populated areas. 

\emph{\textbf{Barab\'{a}si-Albert (BA) graphs}}:
The BA model \cite{BAgraph} provides a random power-law graph generated by the following preferential attachment mechanism. We begin with a small connected graph $\mathcal{G}_0:=(\{v_1,v_2,v_3,v_4\},\{v_1v_2,v_1v_3,v_1v_4\})$ and add nodes sequentially. For each new node $v$, we connect $v$ to $n_{\min}$ existing nodes, where $n_{\min}$ specifies (a lower bound on) the minimum node degree, such that the probability of connecting the new node to existing node $w$ is proportional to the degree of $w$. If the number of existing nodes is smaller than $n_{\min}$, then $v$ connects to all the existing nodes. The BA graph has been used to model many naturally occurring networks, {\clr e.g., citation networks, and social networks.}

\emph{\textbf{Random Power Law (RPL) graphs}}:
The BA model introduces an artifact that all node degrees are lower bounded by $n_{\min}$. Alternatively, the RPL graph \cite{Chung06book} provides another way of generating power-law graphs by directly specifying a sequence of expected node degrees ($d_1,...,d_{|V|}$) according to the power law, i.e., $d_i=i^\alpha$ ($\alpha>0$). The generation of a RPL graph is similar to that of an ER graph, except that instead of connecting each pair of nodes with the same probability, nodes $i$ and $j$ in a RPL graph are connected by a link with probability $p_{ij}=d_id_j/\sum^{|V|}_{k=1}d_k$.

\subsection{Tightness of Bounds}
\label{sect:tightnessBounds}

To measure the impact of probing on the node maximum identifiability $\Omega(v)$ or the maximum identifiable set $S^*(k)$, we need tight bounds under CSP and UP (we can compute the exact value under CAP). Although we have achieved this theoretically by deriving upper and lower bounds, only the bounds under CSP can be evaluated efficiently, and the bounds under UP have to be relaxed by a logarithmic factor to be computable in polynomial time. The first question is therefore how tight the relaxed bounds are.

\begin{figure}[t]
\vspace{-.7em}
\begin{minipage}{.5\linewidth}
  \centerline{\includegraphics[width=1.05\columnwidth]{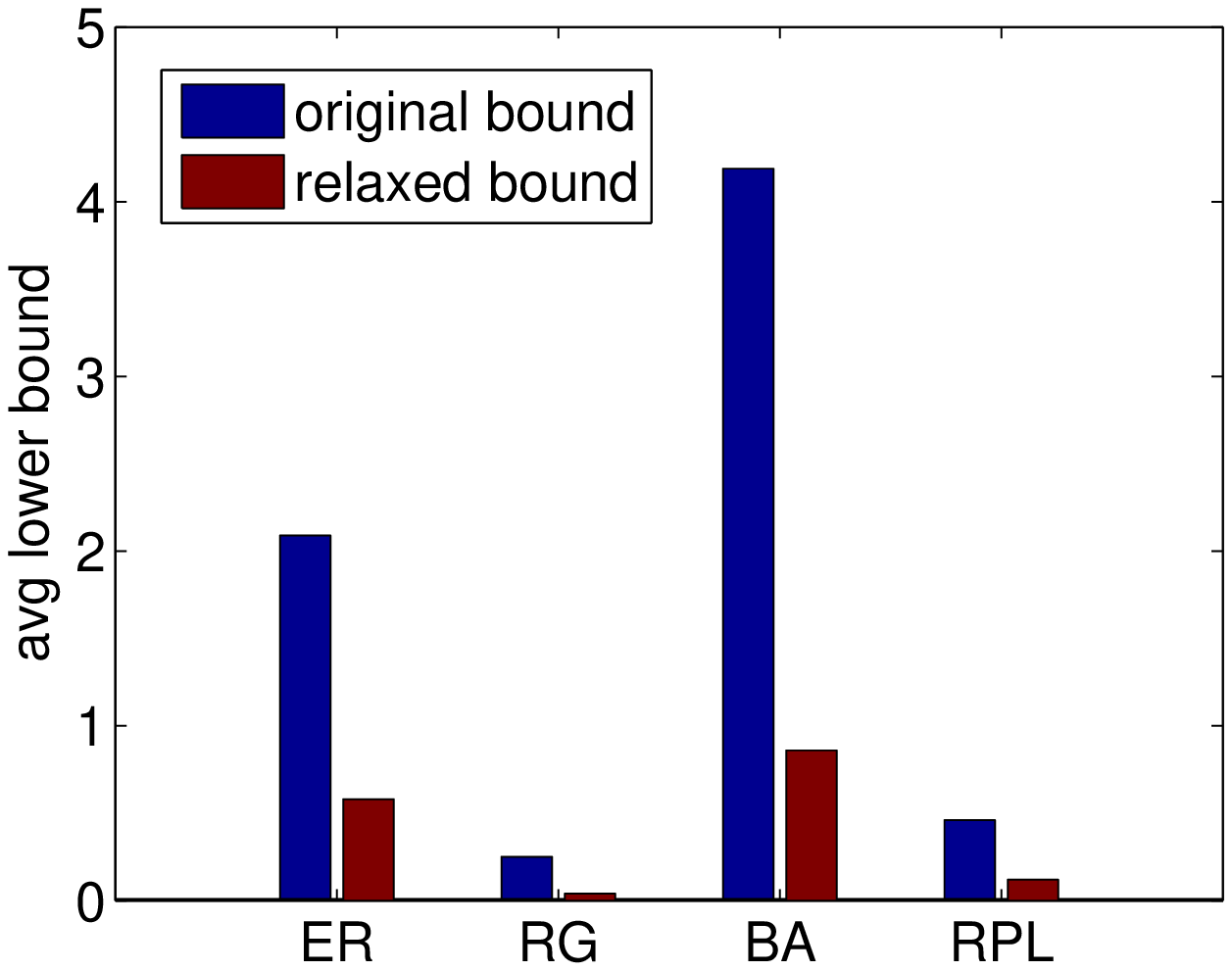}}
  \vspace{-.5em}\centerline{\small (a) Average lower bound}
\end{minipage}\hfill
\begin{minipage}{.5\linewidth}
  \centerline{\includegraphics[width=1.05\columnwidth]{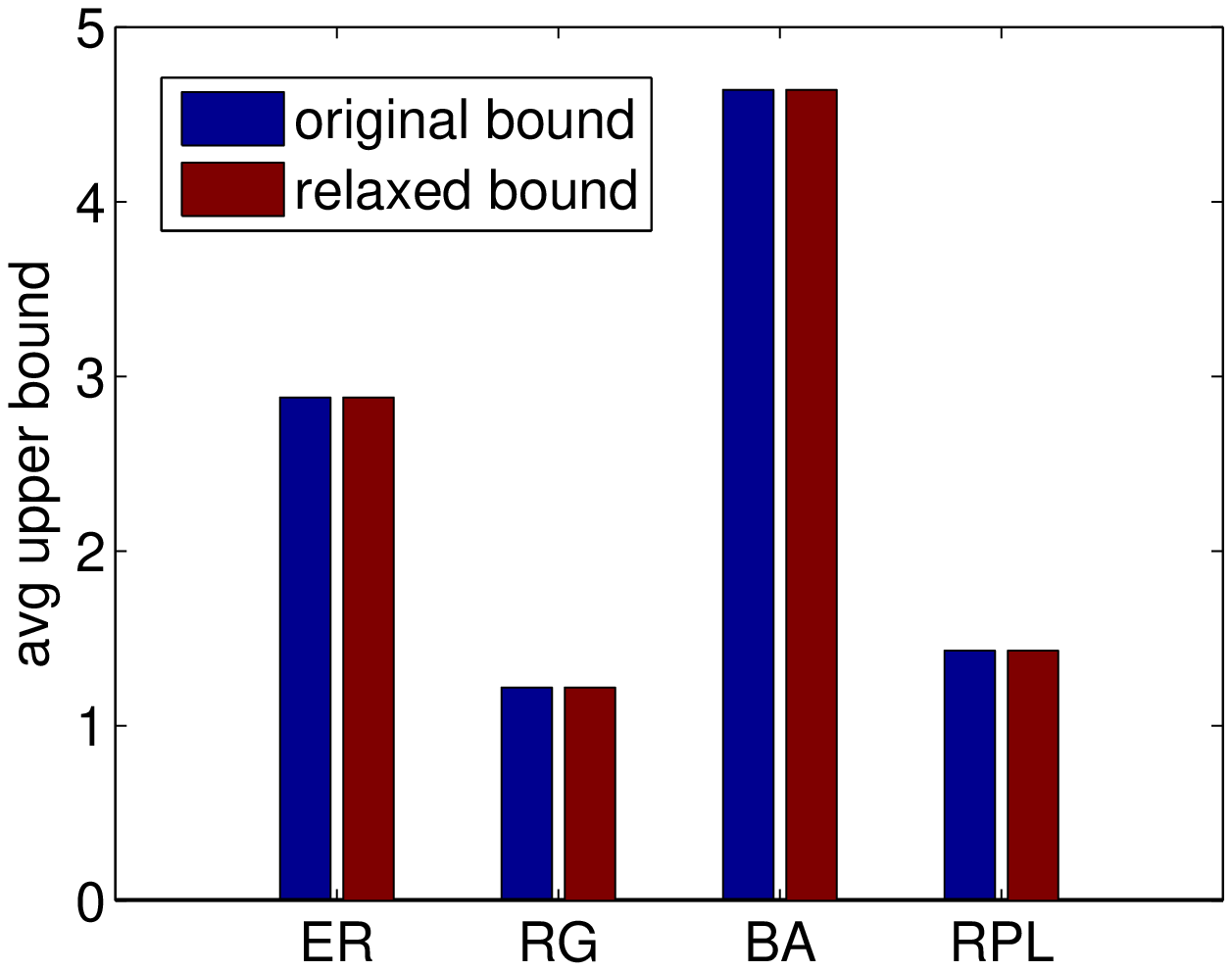}}
  \vspace{-.5em}\centerline{\small (b) Average upper bound}
\end{minipage}
\vspace{-.0em}
\caption{Original and relaxed bounds on the maximum identifiability $\OmegaUP(N)$ under UP for sparsely-connected random topologies ($|V|=20$, $\mu=10$, $\mathbb{E}[|L|]=51$, $100$ graph instances per model). } \label{fig:tightness of Omega_UP bounds}
\end{figure}

\begin{figure}[t]
\vspace{-.5em}
\begin{minipage}{.5\linewidth}
  \centerline{\includegraphics[width=1.05\columnwidth]{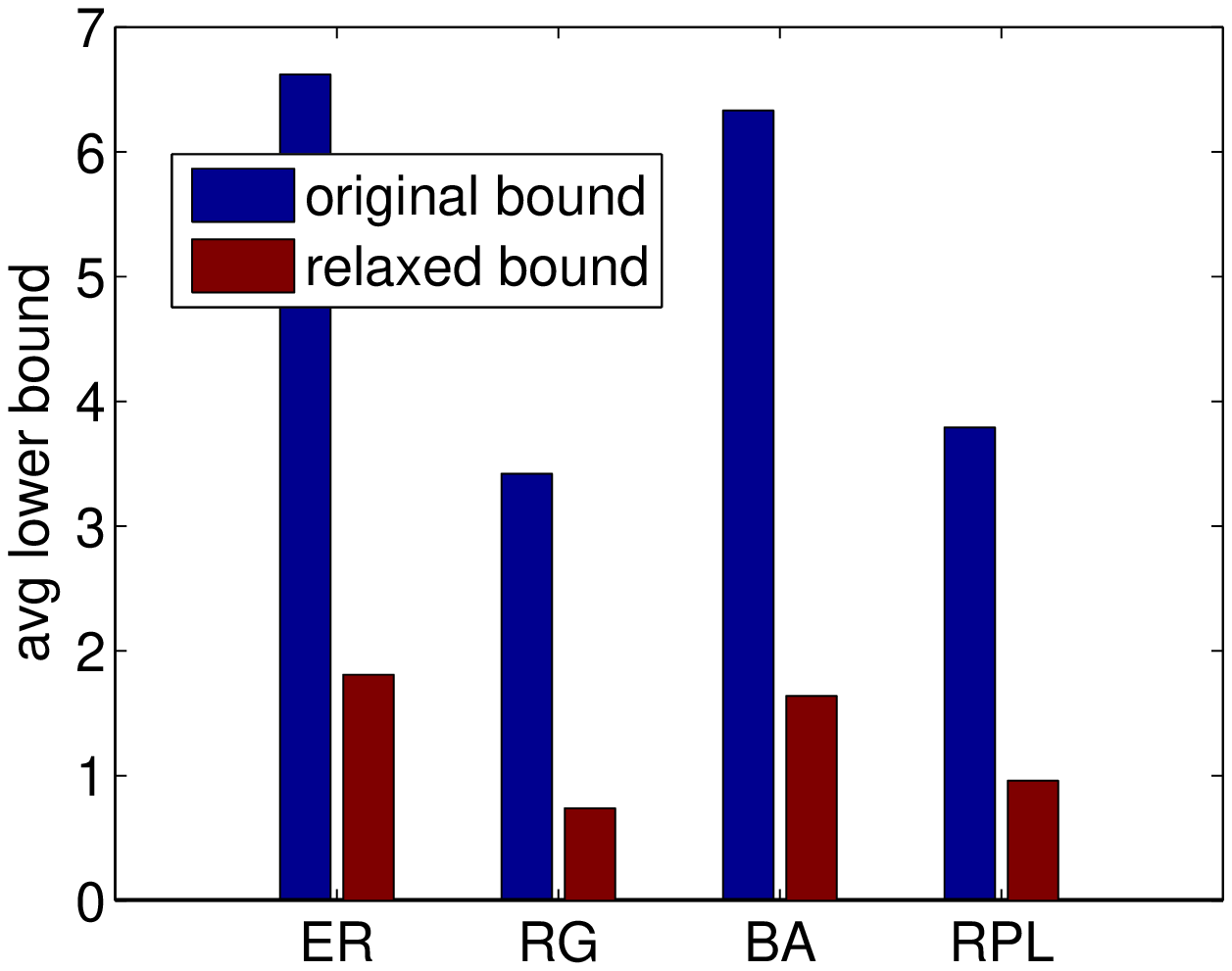}}
  \vspace{-.5em}\centerline{\small (a) Average lower bound}
\end{minipage}\hfill
\begin{minipage}{.5\linewidth}
  \centerline{\includegraphics[width=1.05\columnwidth]{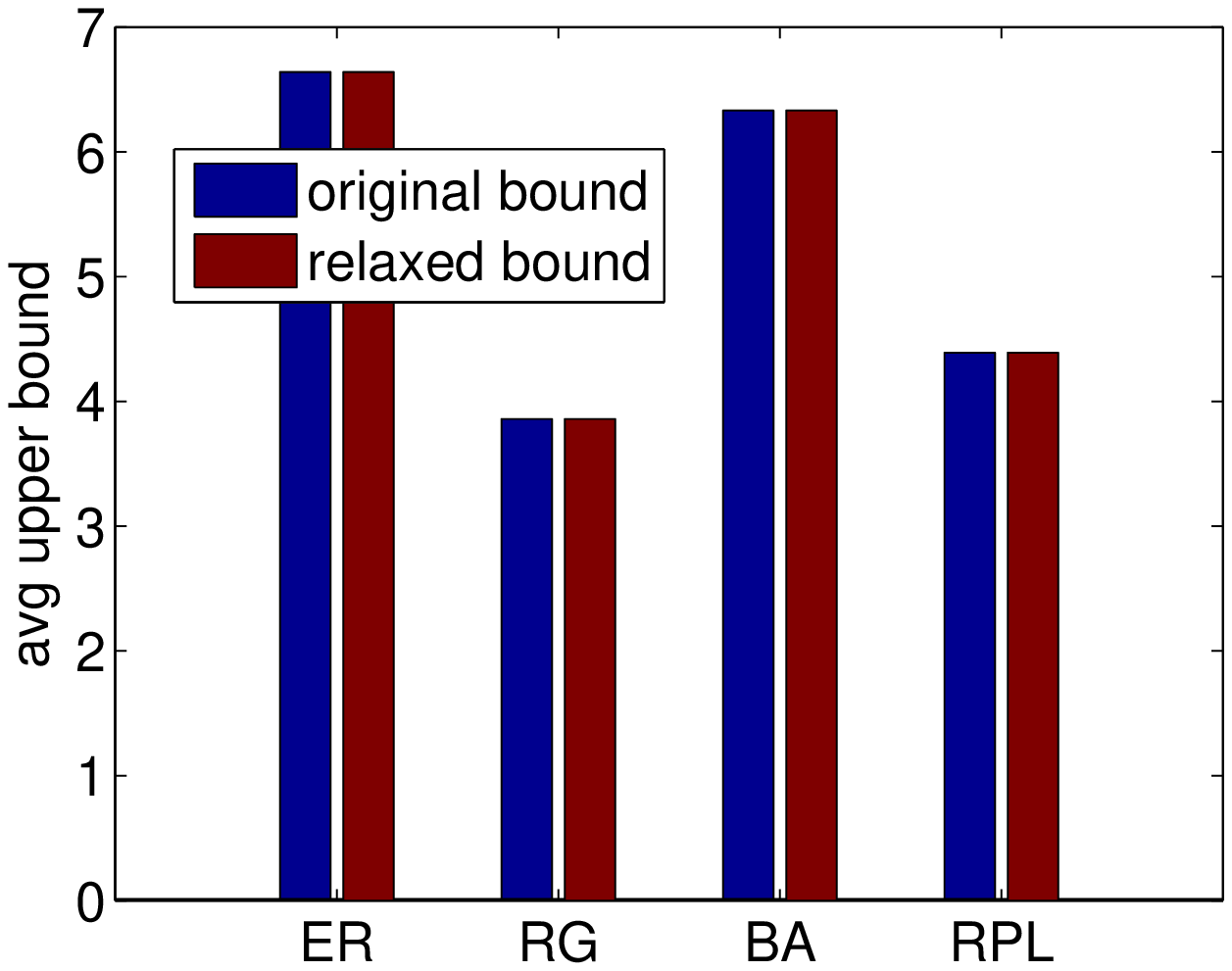}}
  \vspace{-.5em}\centerline{\small (b) Average upper bound}
\end{minipage}
\vspace{-.0em}
\caption{Original and relaxed bounds on the maximum identifiability $\OmegaUP(N)$ under UP for densely-connected random topologies ($|V|=20$, $\mu=10$, $\mathbb{E}[|L|]=99$, $100$ graph instances per model). } \label{fig:tightness of Omega_UP bounds_dense}
\end{figure}

To this end, we compare the original bounds (Theorem~\ref{thm:Omega_UP}) and the relaxed bounds on a variety of topologies synthetically generated from the models in Section~\ref{subsubsec:Random Topologies} in two scenarios, i.e., sparsely-connected and densely-connected topologies. To make the models comparable in each scenario, we have tuned each model to generate graphs with the same number of nodes and (average) number of links. We select $S=N$ (the identification of all non-monitors are of interest), evaluate both bounds on multiple graph instances per model, each with a fixed number of randomly placed monitors, and present the average lower/upper bounds in Fig.~\ref{fig:tightness of Omega_UP bounds} and Fig.~\ref{fig:tightness of Omega_UP bounds_dense}. As expected, in both scenarios, the relaxed lower bounds are quite loose due to the logarithmic factor, but the relaxed upper bounds coincide with the original bounds for all graph instances. This indicates that although $\GSC(v)$ can be a logarithmic-factor larger than the original $\MSC(v)$ in the worst case, this worst case rarely occurs, and we can approximate $\MSC(v)$ by $\GSC(v)$. This provides a tight characterization of $\OmegaUP$ and $\SUP$ for large networks, where computing the original bounds is infeasible.

\subsection{Simulation Results}

\begin{figure}[tb]
\vspace{-.7em}
\begin{minipage}{.5\linewidth}
  \centerline{\includegraphics[width=1.05\columnwidth]{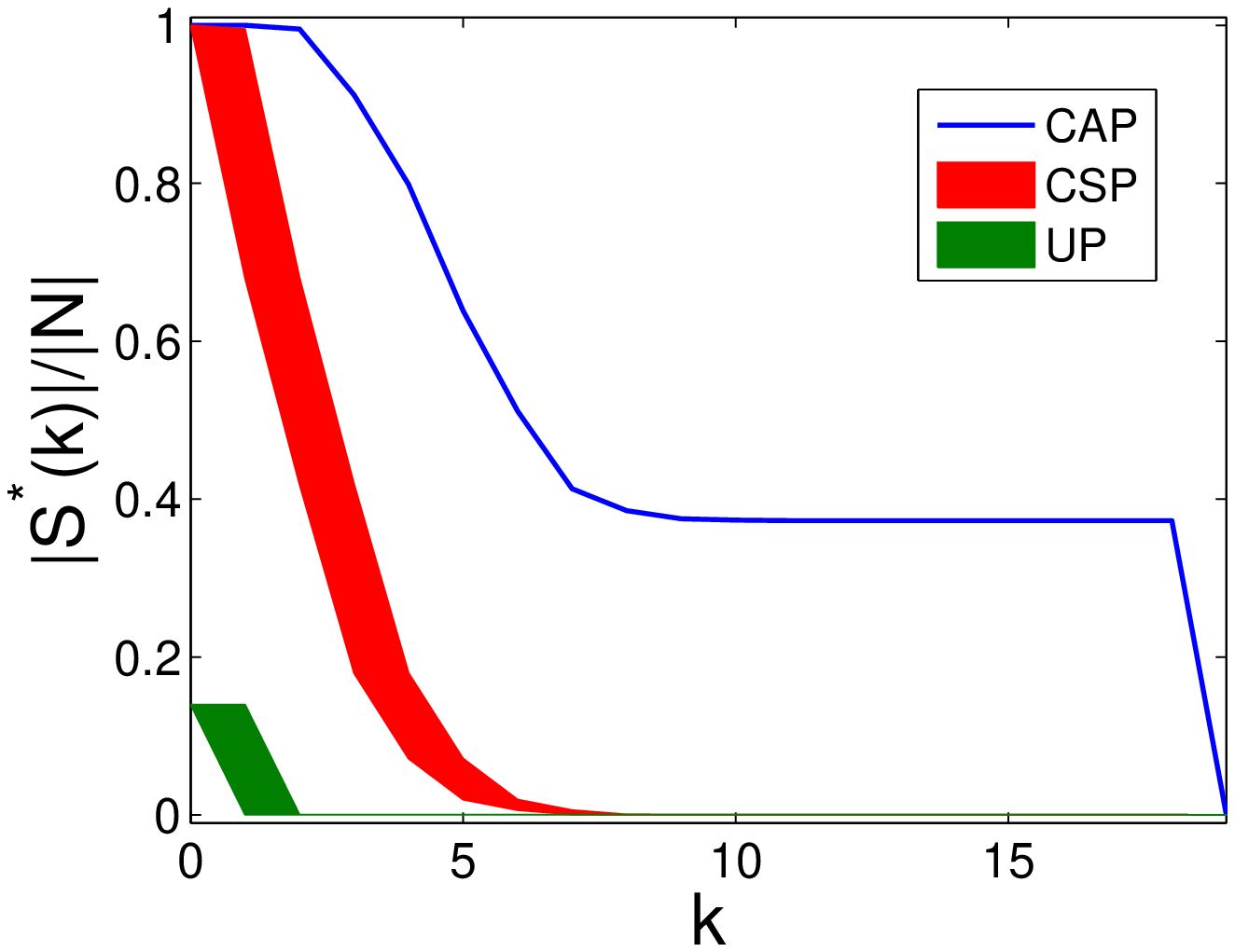}}
  \vspace{-.5em}\centerline{\scriptsize (a) $\mu=2$}
\end{minipage}\hfill
\begin{minipage}{.5\linewidth}
  \centerline{\includegraphics[width=1.05\columnwidth]{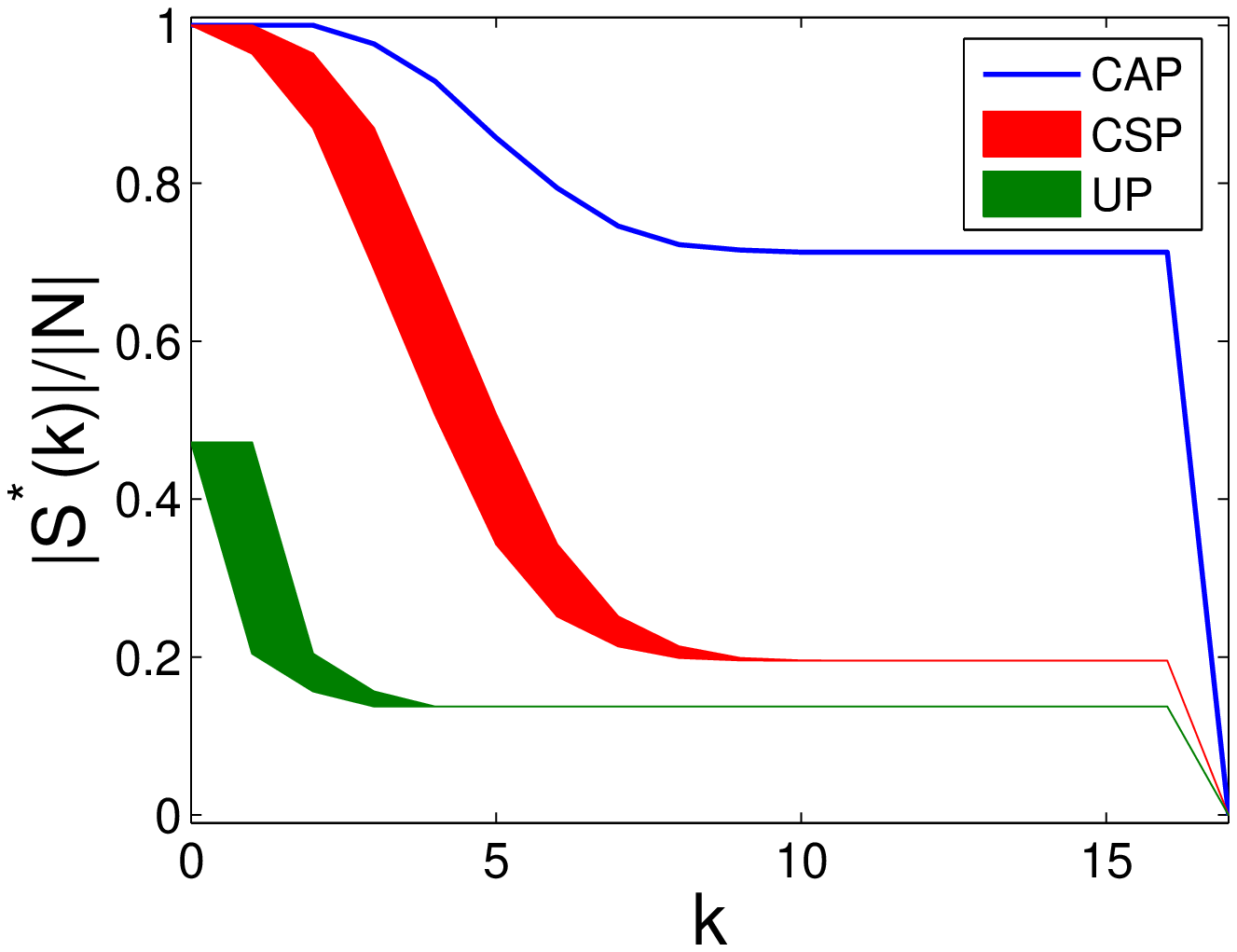}}
  \vspace{-.5em}\centerline{\scriptsize (b) $\mu=4$}
\end{minipage}
\vspace{-.5em}
\begin{minipage}{.5\linewidth}
  \centerline{\includegraphics[width=1.05\columnwidth]{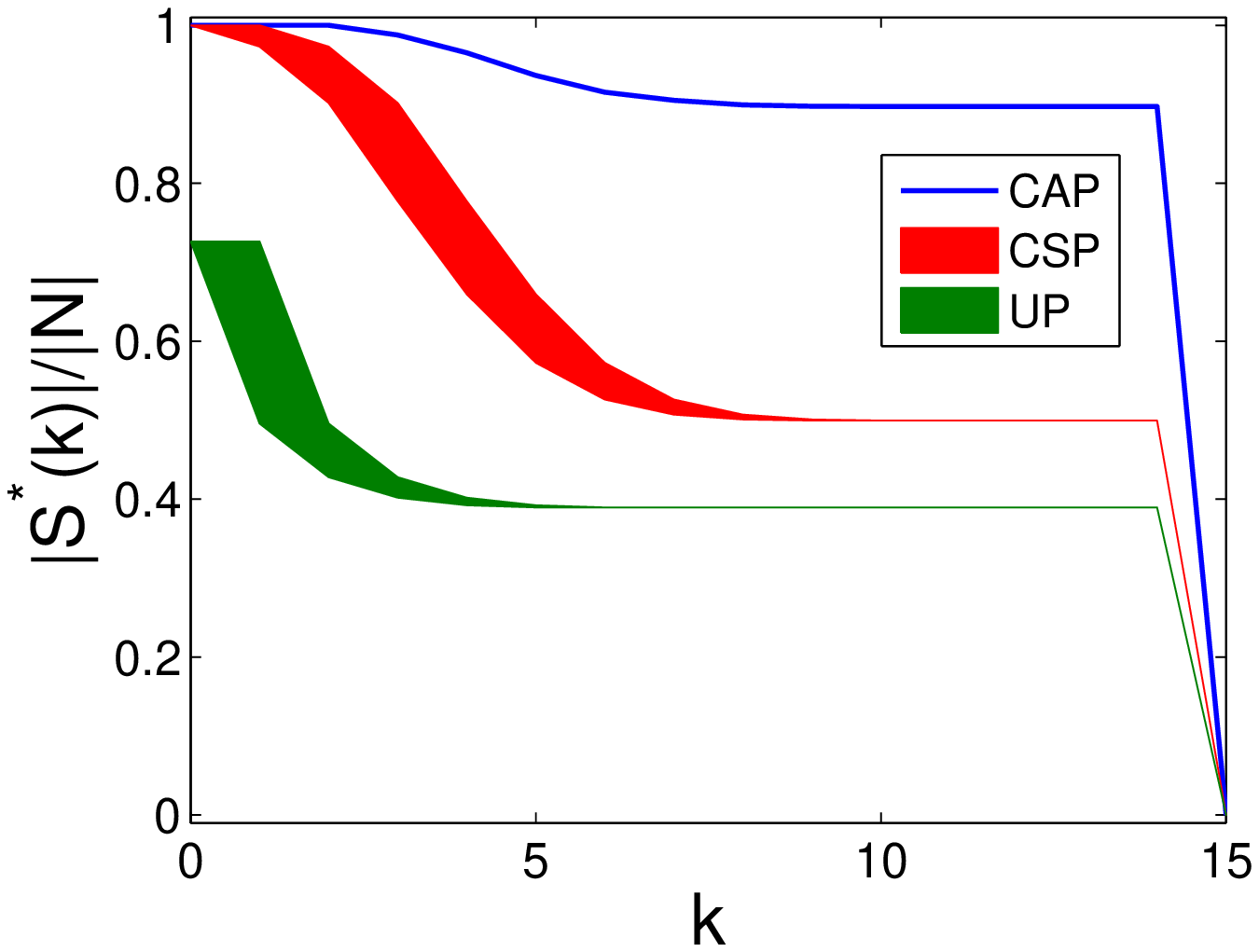}}
  \vspace{-.5em}\centerline{\scriptsize (c) $\mu=6$}
\end{minipage}\hfill
\begin{minipage}{.5\linewidth}
  \centerline{\includegraphics[width=1.05\columnwidth]{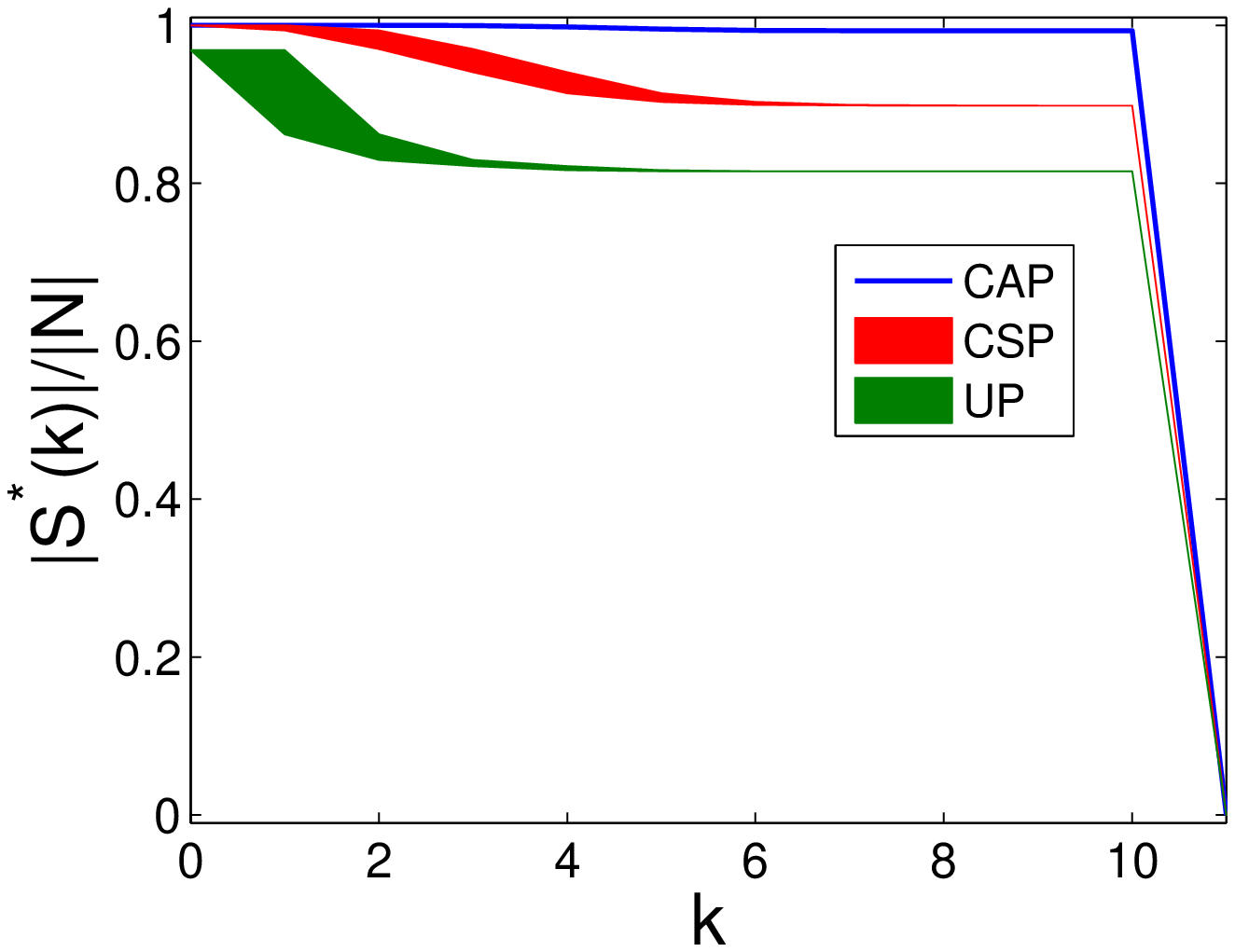}}
  \vspace{-.5em}\centerline{\scriptsize (d) $\mu=10$}
\end{minipage}
\vspace{-.0em}
\caption{Maximum $k$-identifiable set $S^*(k)$ under CAP, CSP, and UP for ER graphs ($|V|=20$, $\mu=\{2,4,6,10\}$, $\mathbb{E}[|L|]=51$, $200$ graph instances). } \label{fig:S_bounds_ER}
\end{figure}

\begin{figure}[tb]
\vspace{-.7em}
\begin{minipage}{.5\linewidth}
  \centerline{\includegraphics[width=1.05\columnwidth]{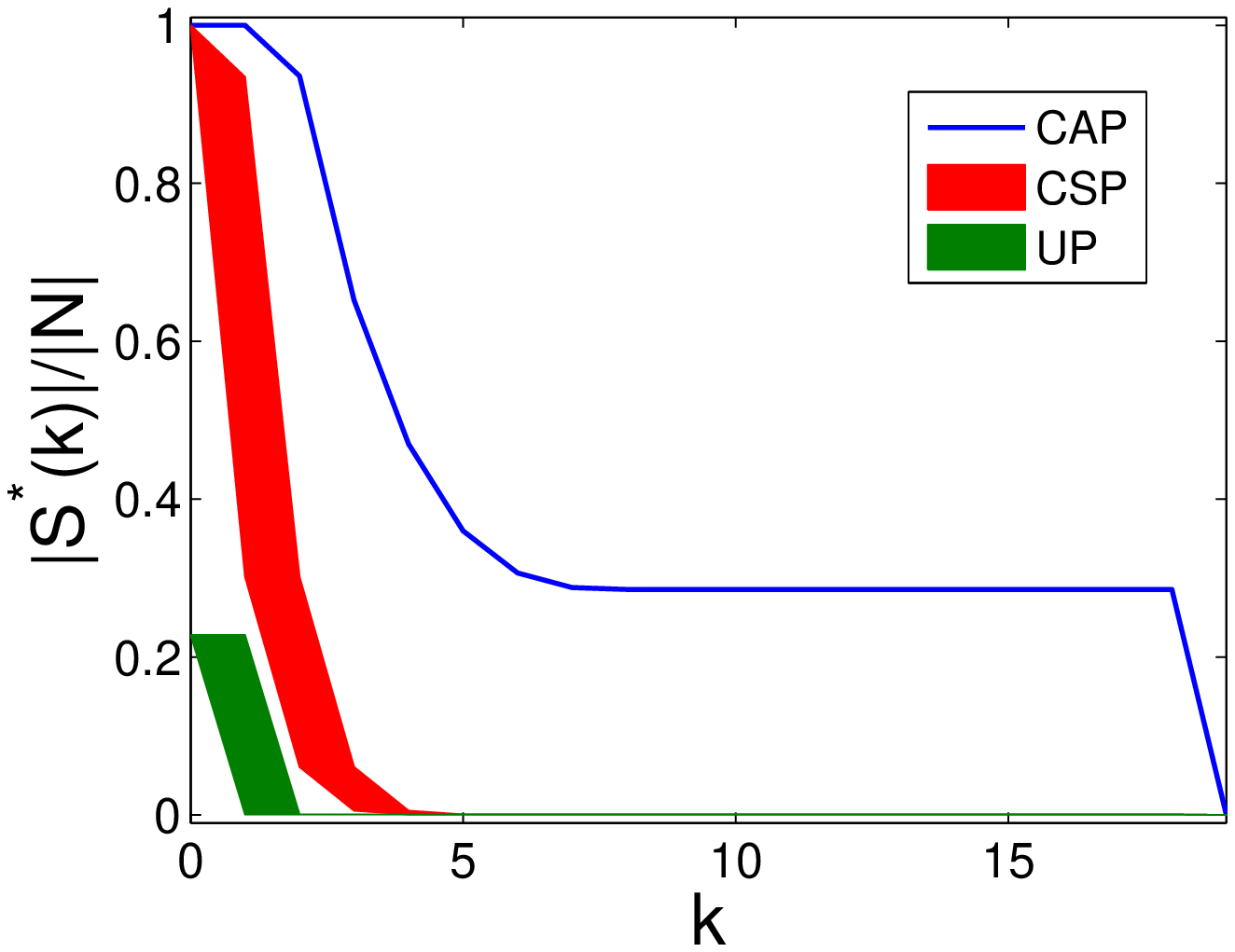}}
  \vspace{-.5em}\centerline{\scriptsize (a) $\mu=2$}
\end{minipage}\hfill
\begin{minipage}{.5\linewidth}
  \centerline{\includegraphics[width=1.05\columnwidth]{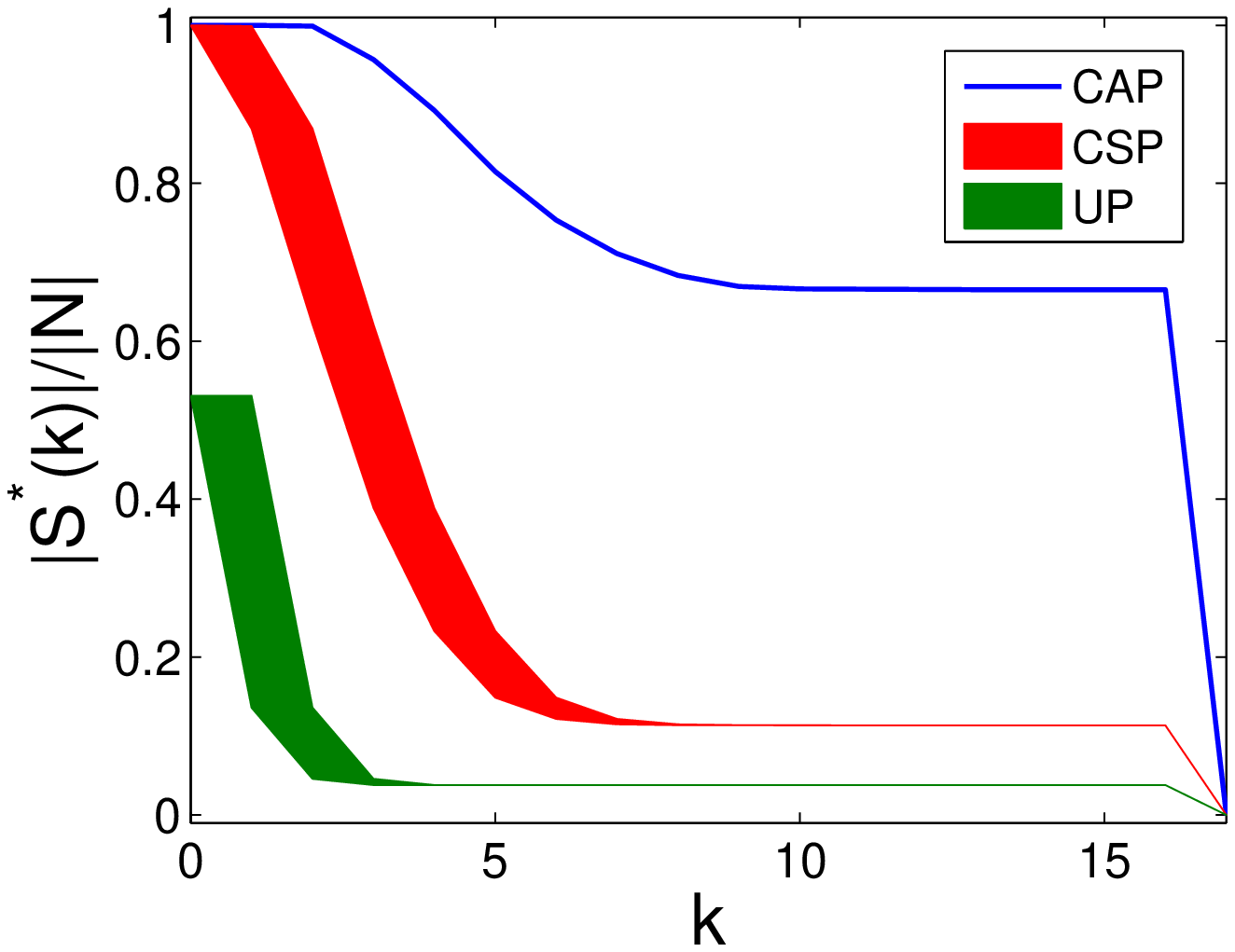}}
  \vspace{-.5em}\centerline{\scriptsize (b) $\mu=4$}
\end{minipage}
\vspace{-.5em}
\begin{minipage}{.5\linewidth}
  \centerline{\includegraphics[width=1.05\columnwidth]{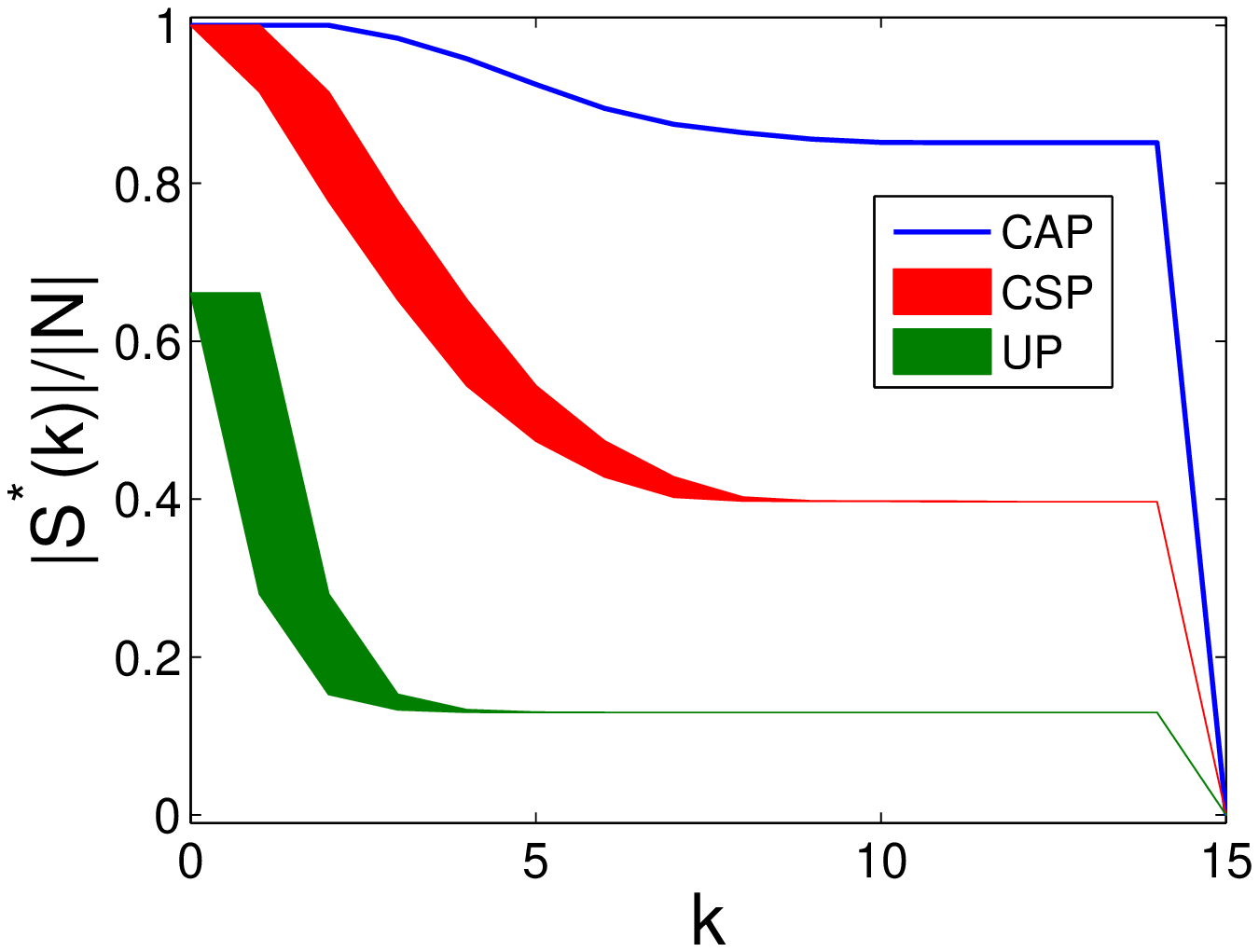}}
  \vspace{-.5em}\centerline{\scriptsize (c) $\mu=6$}
\end{minipage}\hfill
\begin{minipage}{.5\linewidth}
  \centerline{\includegraphics[width=1.05\columnwidth]{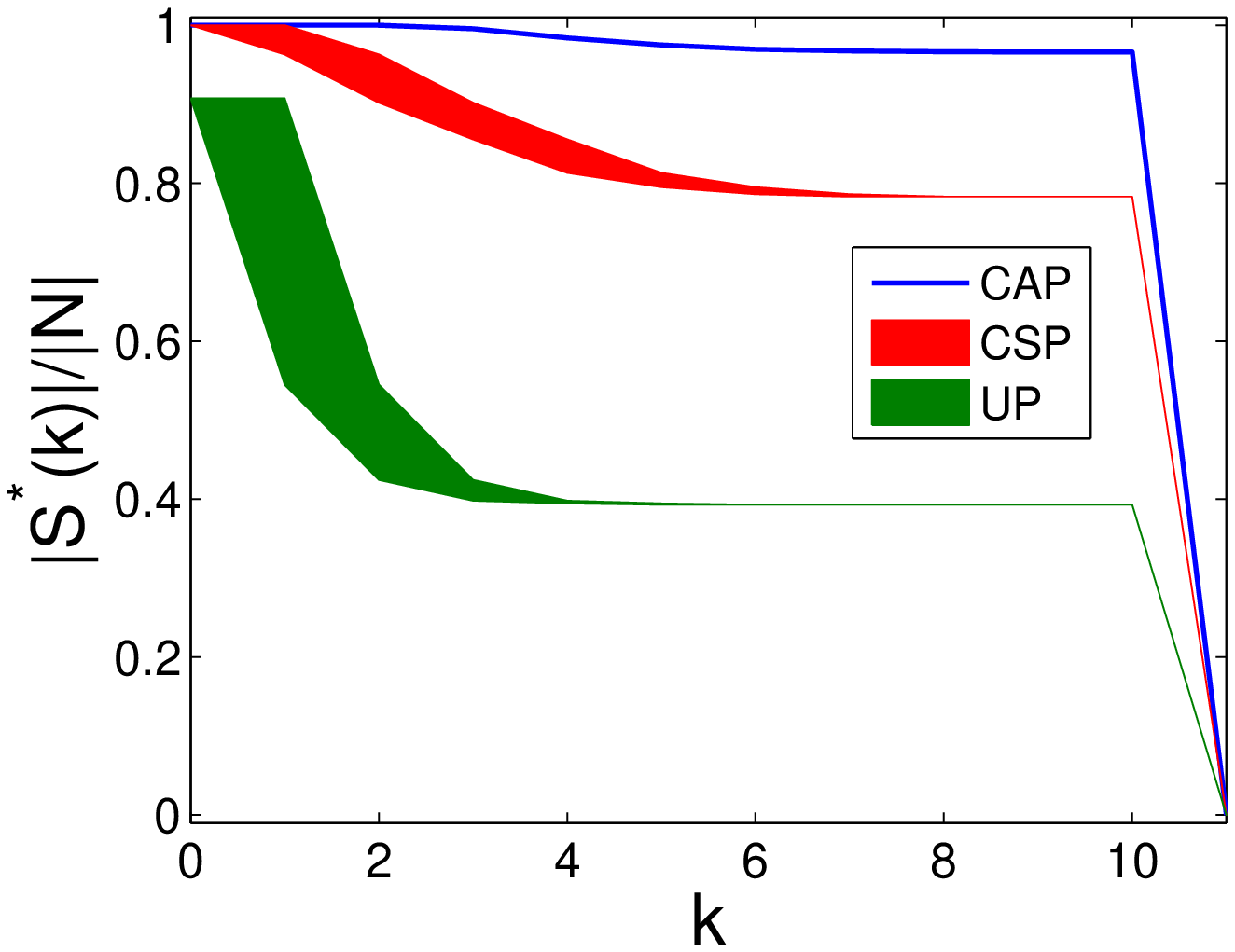}}
  \vspace{-.5em}\centerline{\scriptsize (d) $\mu=10$}
\end{minipage}
\vspace{-.0em}
\caption{Maximum $k$-identifiable set $S^*(k)$ under CAP, CSP, and UP for RG graphs ($|V|=20$, $\mu=\{2,4,6,10\}$, $\mathbb{E}[|L|]=51$, $200$ graph instances). } \label{fig:S_bounds_RG}
\end{figure}

\begin{figure}[tb]
\vspace{-.7em}
\begin{minipage}{.5\linewidth}
  \centerline{\includegraphics[width=1.05\columnwidth]{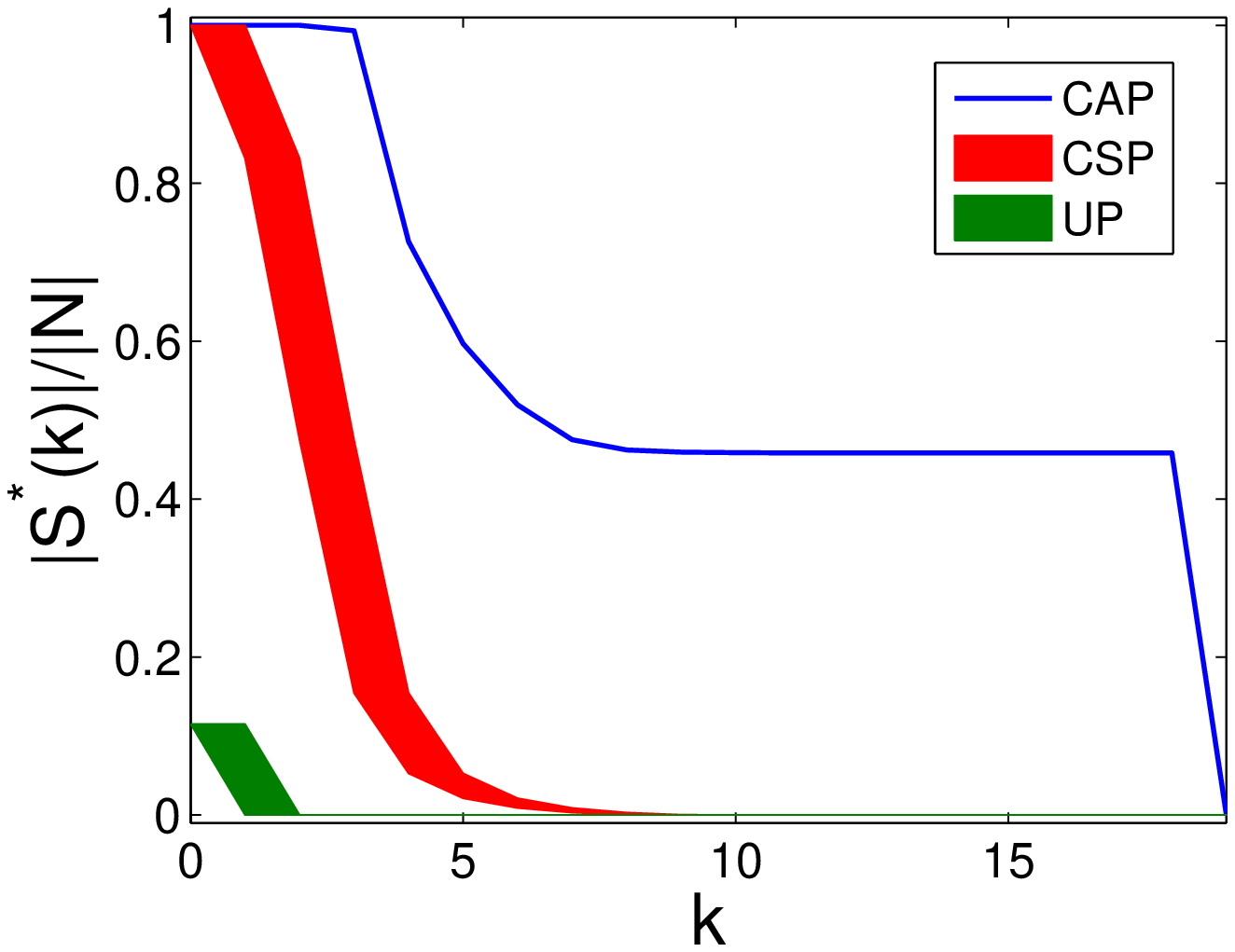}}
  \vspace{-.5em}\centerline{\scriptsize (a) $\mu=2$}
\end{minipage}\hfill
\begin{minipage}{.5\linewidth}
  \centerline{\includegraphics[width=1.05\columnwidth]{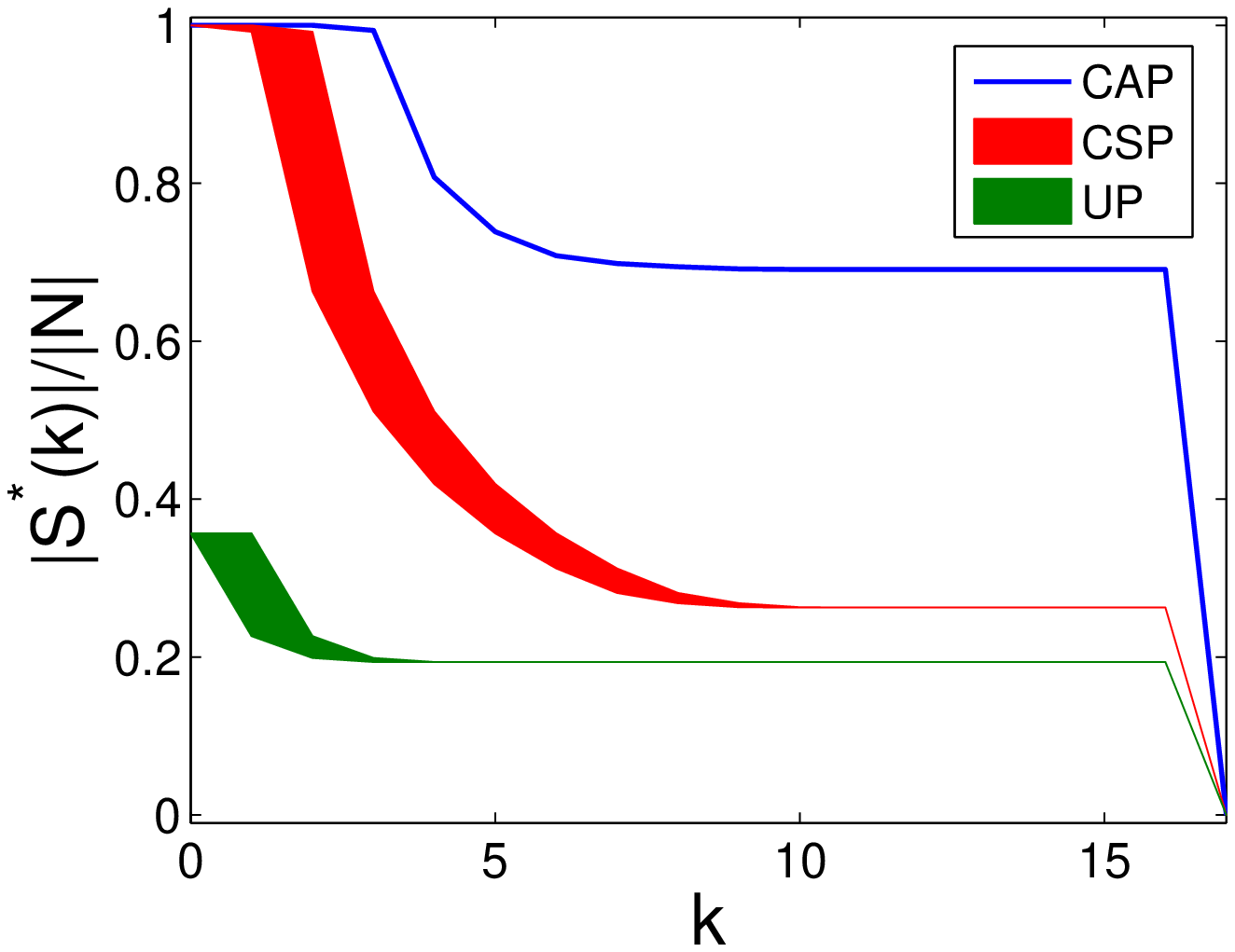}}
  \vspace{-.5em}\centerline{\scriptsize (b) $\mu=4$}
\end{minipage}
\vspace{-.5em}
\begin{minipage}{.5\linewidth}
  \centerline{\includegraphics[width=1.05\columnwidth]{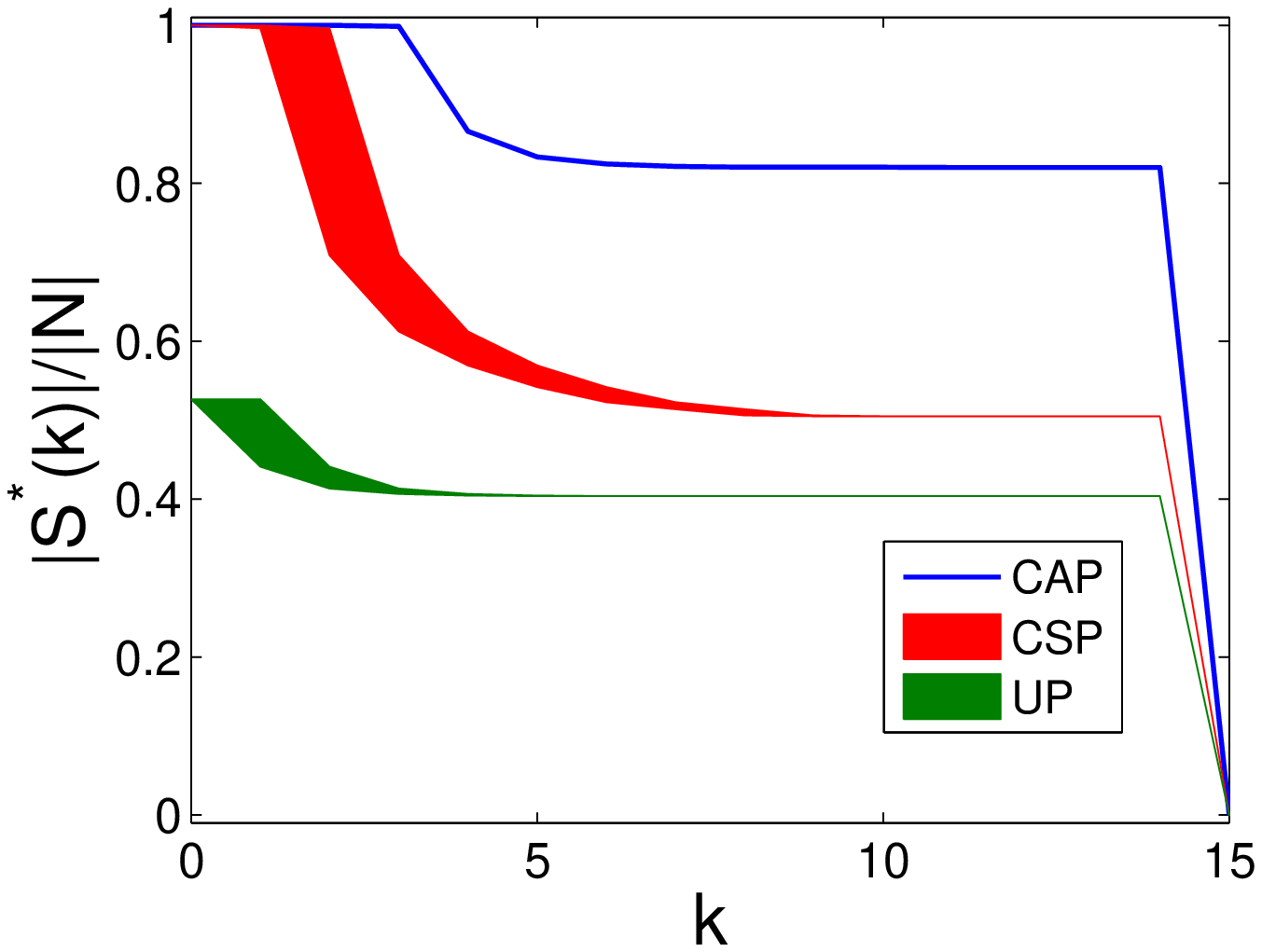}}
  \vspace{-.5em}\centerline{\scriptsize (c) $\mu=6$}
\end{minipage}\hfill
\begin{minipage}{.5\linewidth}
  \centerline{\includegraphics[width=1.05\columnwidth]{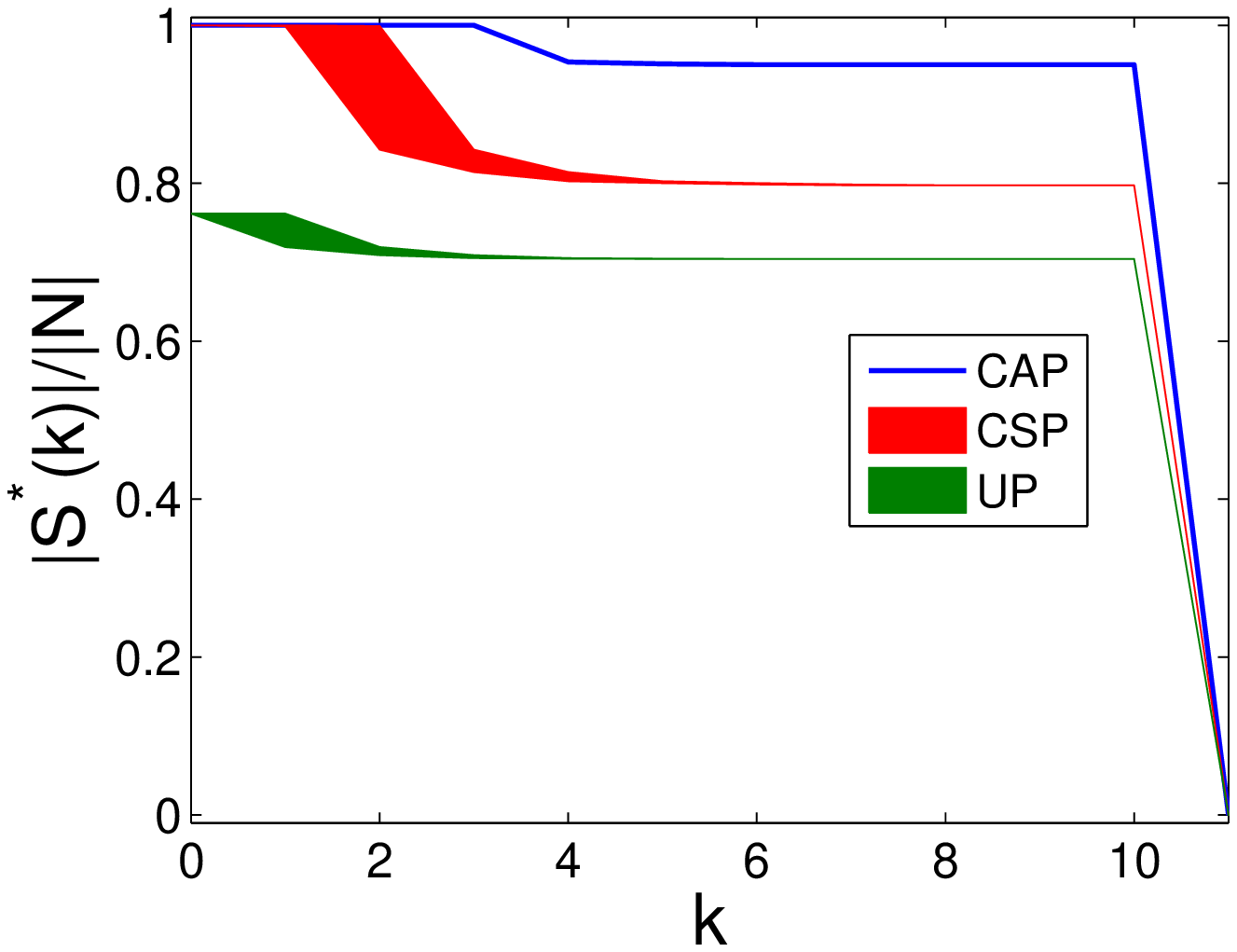}}
  \vspace{-.5em}\centerline{\scriptsize (d) $\mu=10$}
\end{minipage}
\vspace{-.0em}
\caption{Maximum $k$-identifiable set $S^*(k)$ under CAP, CSP, and UP for BA graphs ($|V|=20$, $\mu=\{2,4,6,10\}$, $\mathbb{E}[|L|]=51$, $200$ graph instances). } \label{fig:S_bounds_BA}
\end{figure}

\begin{figure}[tb]
\vspace{-.7em}
\begin{minipage}{.5\linewidth}
  \centerline{\includegraphics[width=1.05\columnwidth]{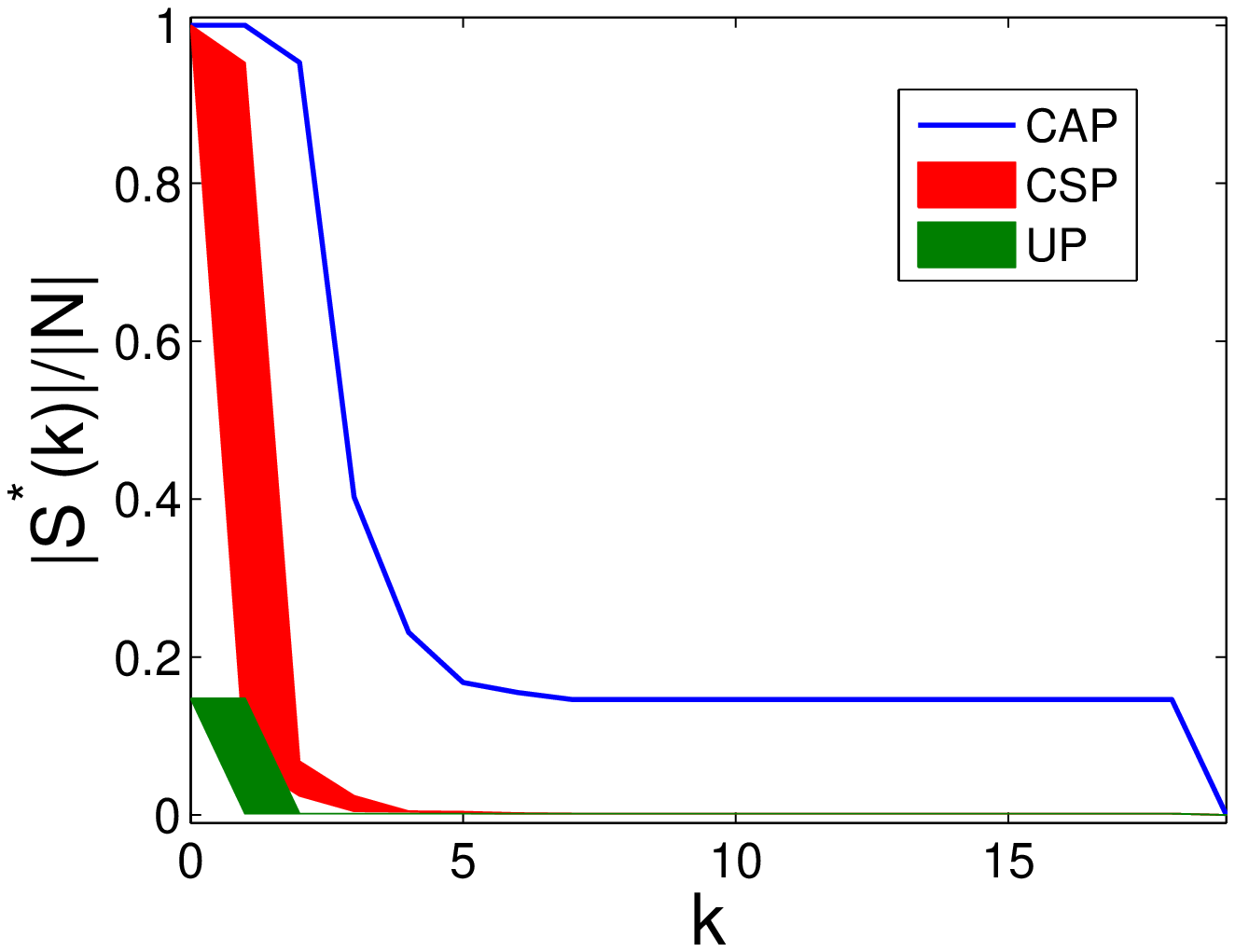}}
  \vspace{-.5em}\centerline{\scriptsize (a) $\mu=2$}
\end{minipage}\hfill
\begin{minipage}{.5\linewidth}
  \centerline{\includegraphics[width=1.05\columnwidth]{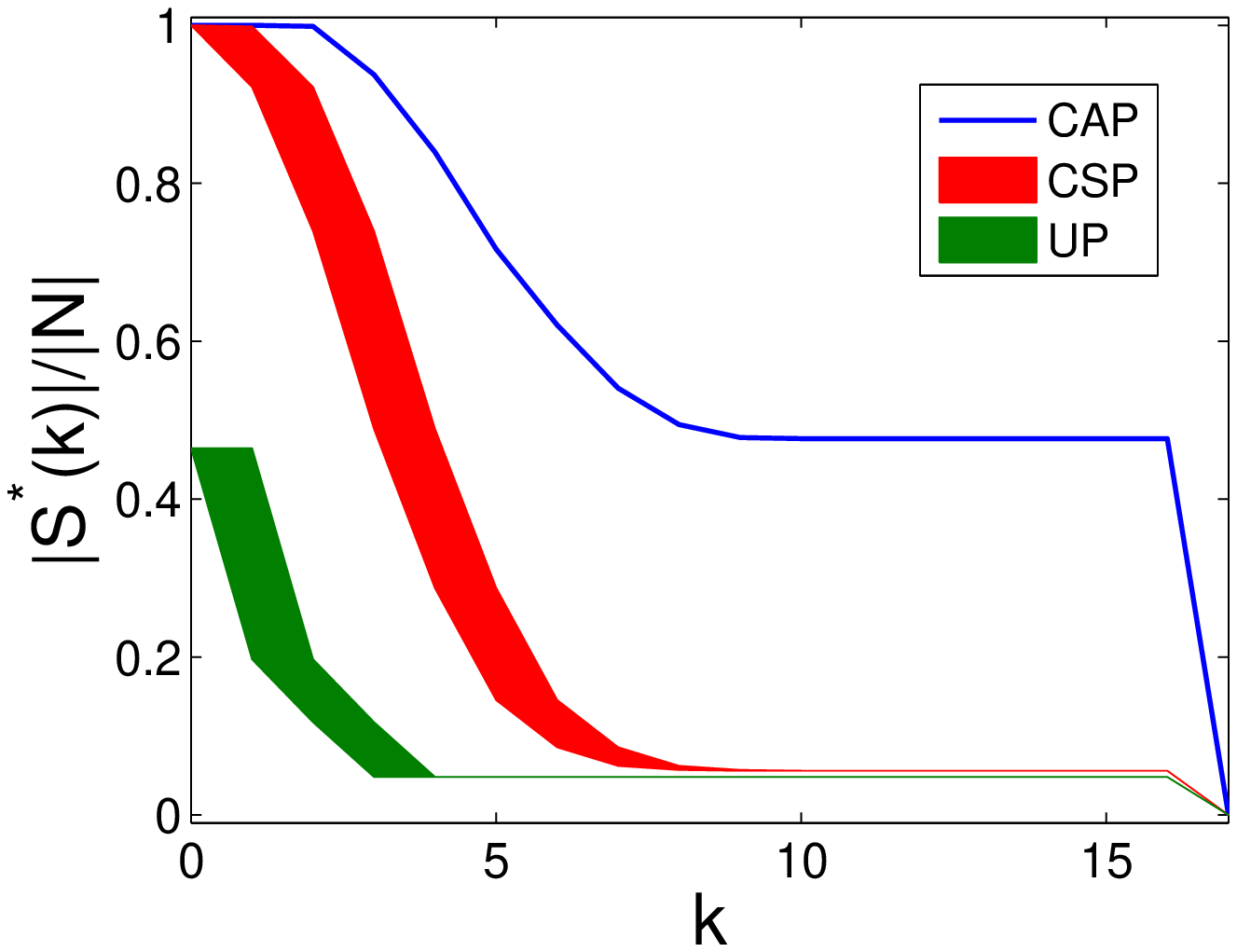}}
  \vspace{-.5em}\centerline{\scriptsize (b) $\mu=4$}
\end{minipage}
\vspace{-.5em}
\begin{minipage}{.5\linewidth}
  \centerline{\includegraphics[width=1.05\columnwidth]{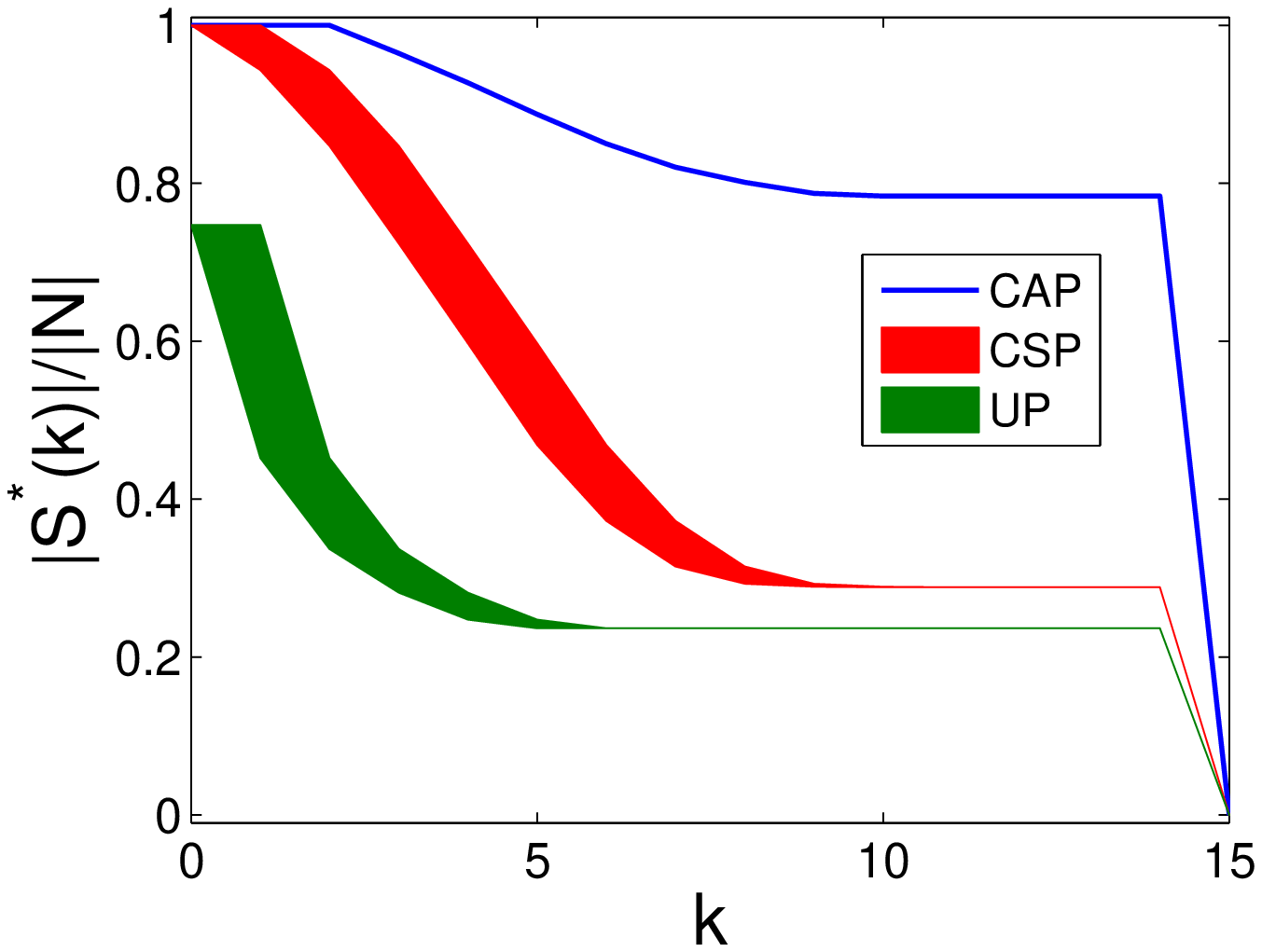}}
  \vspace{-.5em}\centerline{\scriptsize (c) $\mu=6$}
\end{minipage}\hfill
\begin{minipage}{.5\linewidth}
  \centerline{\includegraphics[width=1.05\columnwidth]{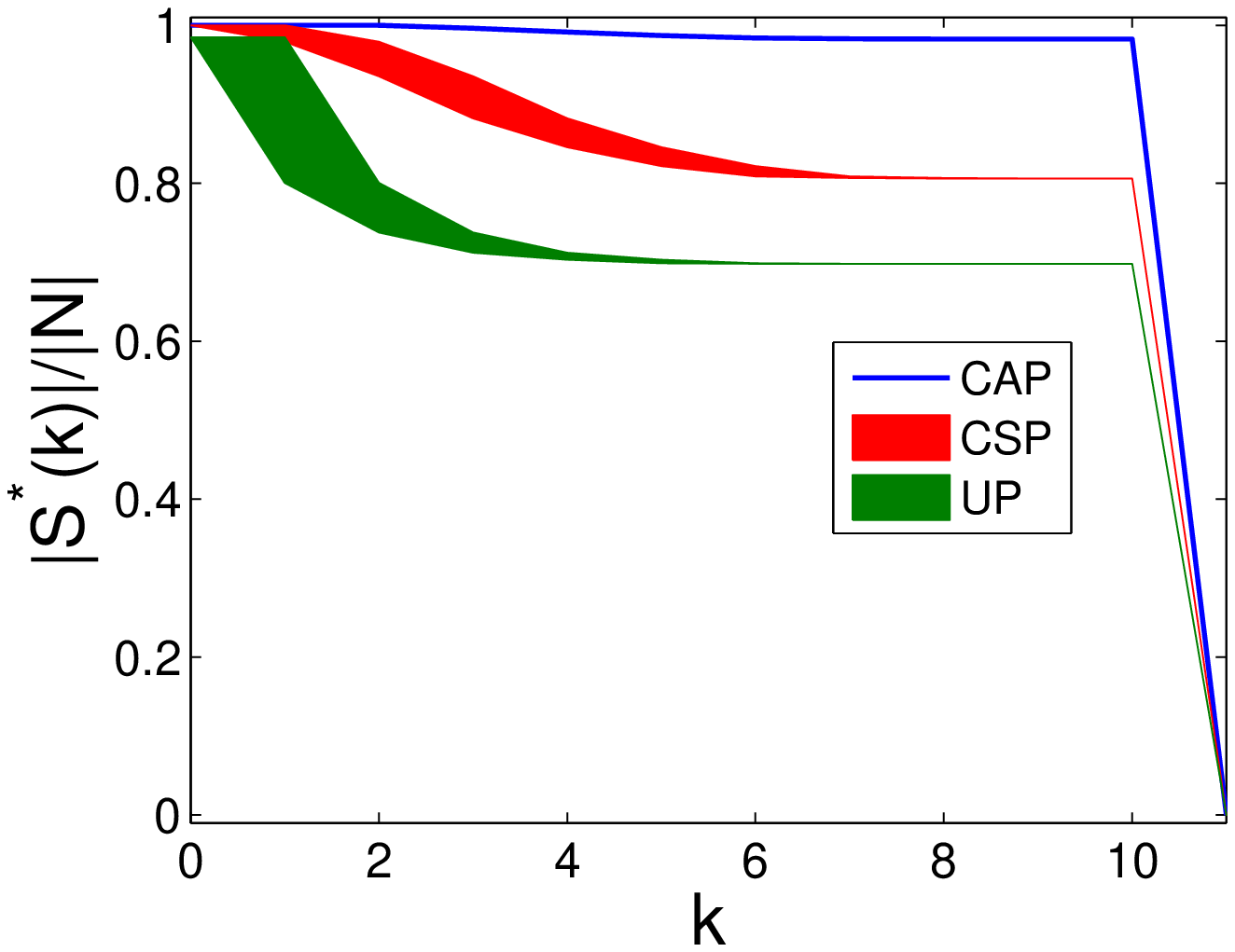}}
  \vspace{-.5em}\centerline{\scriptsize (d) $\mu=10$}
\end{minipage}
\vspace{-.0em}
\caption{Maximum $k$-identifiable set $S^*(k)$ under CAP, CSP, and UP for RPL graphs ($|V|=20$, $\mu=\{2,4,6,10\}$, $\mathbb{E}[|L|]=51$, $200$ graph instances). } \label{fig:S_bounds_RPL}
\end{figure}

\begin{figure}[tb]
\vspace{-.7em}
\begin{minipage}{.5\linewidth}
  \centerline{\includegraphics[width=1.05\columnwidth]{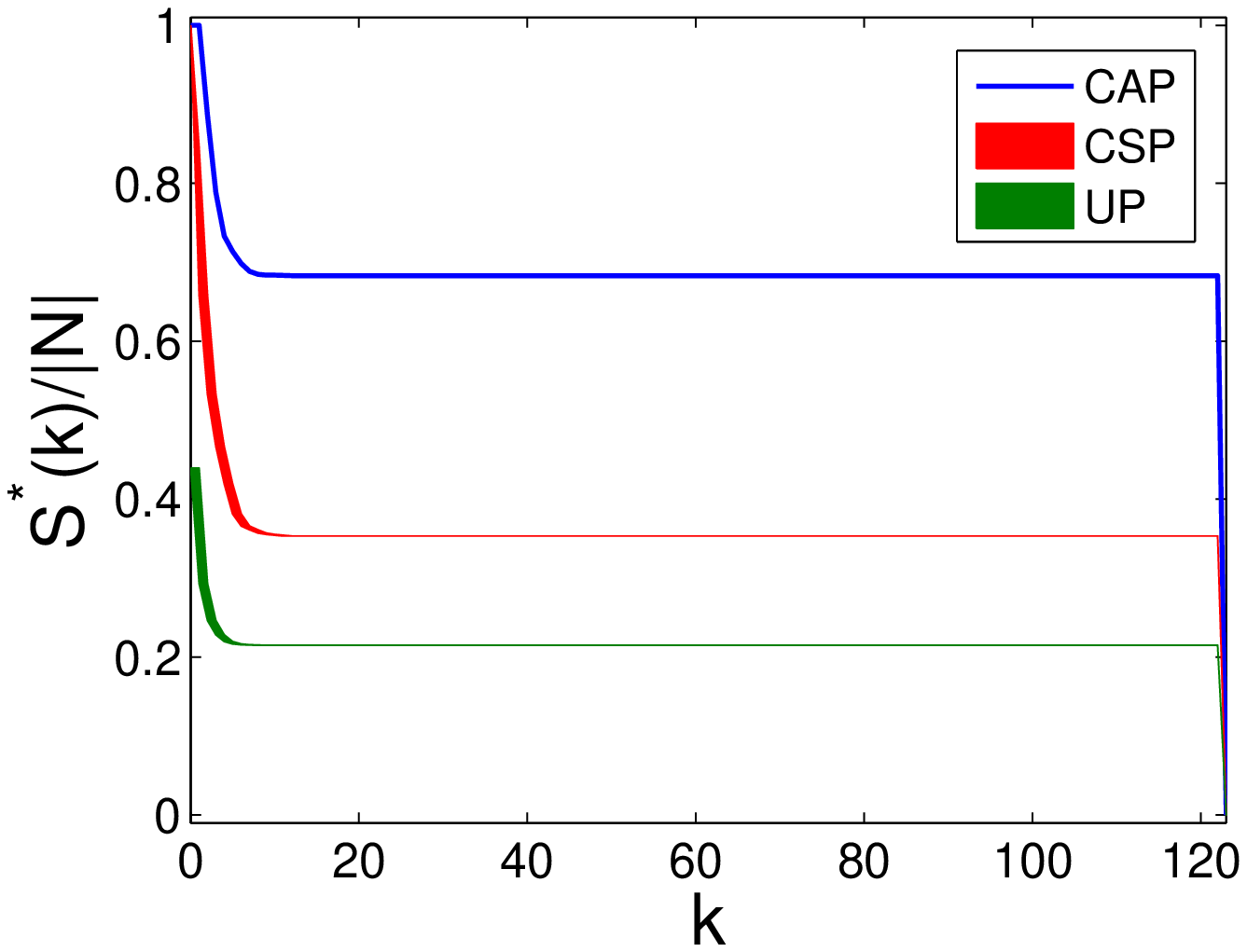}}
  \vspace{-.5em}\centerline{\scriptsize (a) $\mu=50$}
\end{minipage}\hfill
\begin{minipage}{.5\linewidth}
  \centerline{\includegraphics[width=1.05\columnwidth]{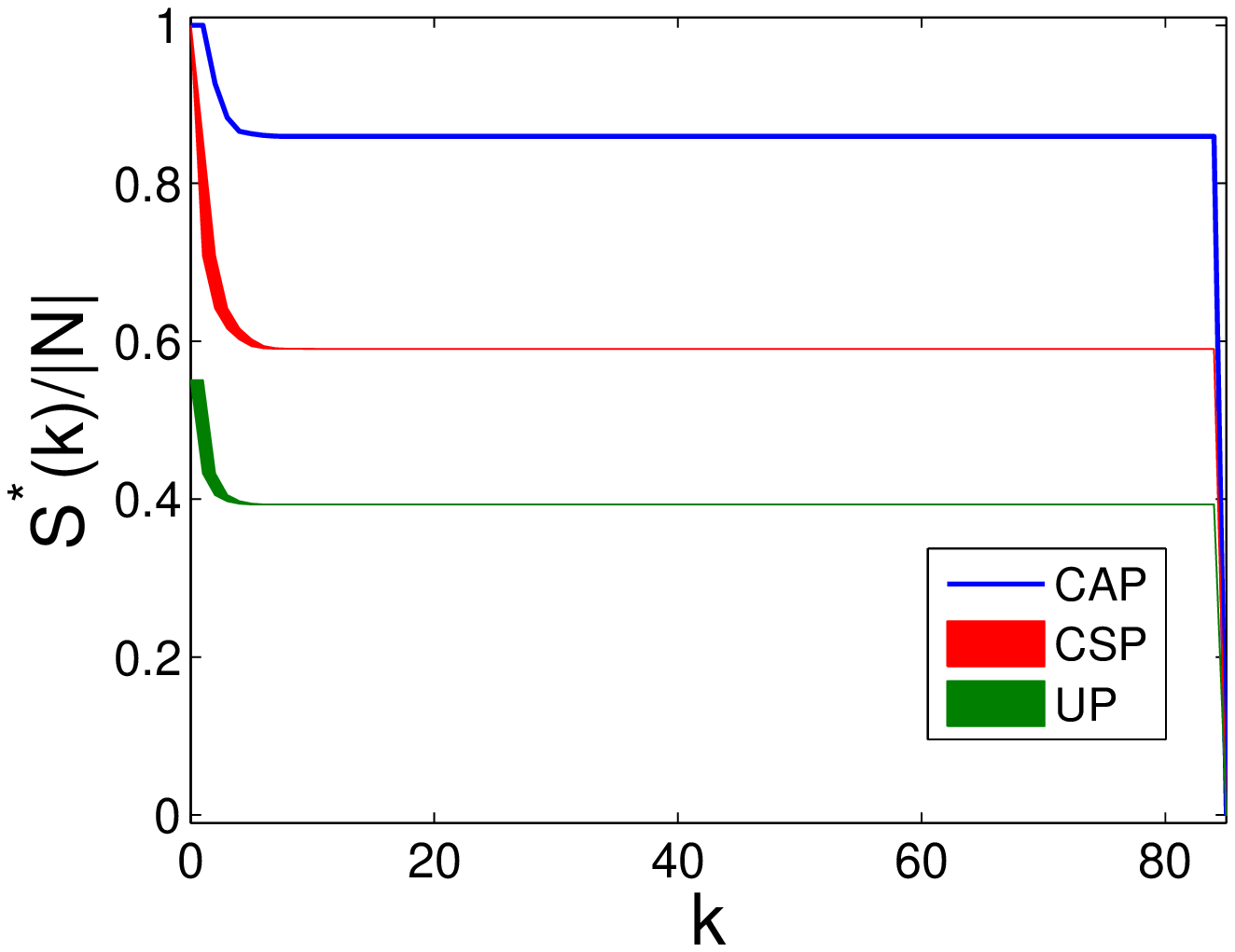}}
  \vspace{-.5em}\centerline{\scriptsize (b) $\mu=88$}
\end{minipage}
\vspace{-.5em}
\begin{minipage}{.5\linewidth}
  \centerline{\includegraphics[width=1.05\columnwidth]{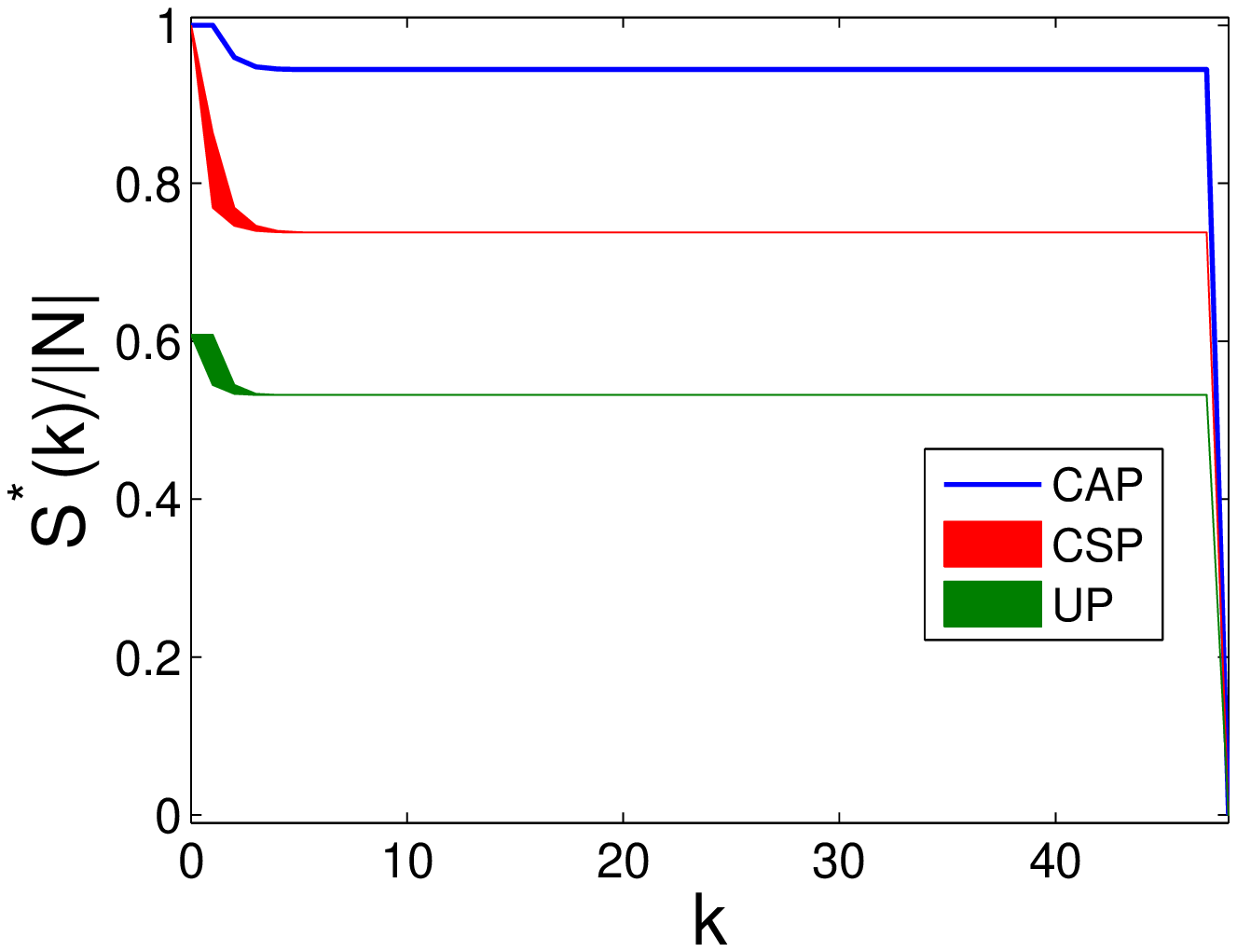}}
  \vspace{-.5em}\centerline{\scriptsize (c) $\mu=125$}
\end{minipage}\hfill
\begin{minipage}{.5\linewidth}
  \centerline{\includegraphics[width=1.05\columnwidth]{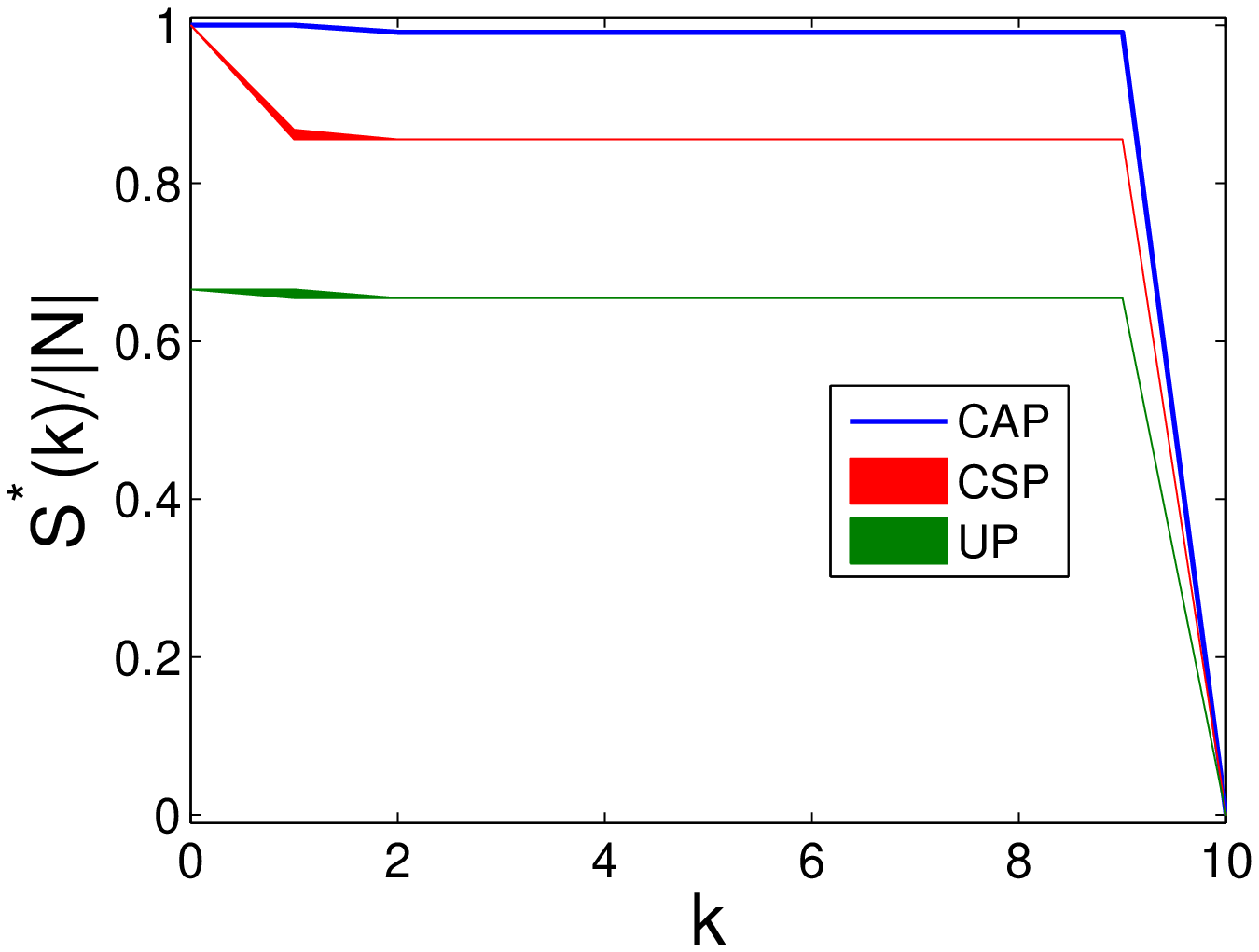}}
  \vspace{-.5em}\centerline{\scriptsize (d) $\mu=163$}
\end{minipage}
\vspace{-.0em}
\caption{Maximum $k$-identifiable set $S^*(k)$ under CAP, CSP, and UP for Rocketfuel AS1755 ($|V|=172$, $|L|=381$, $\mu=\{50,88,125,163\}$, $100$ Monte Carlo runs). } \label{fig:S_bounds_Rocketfuel}
\end{figure}

\begin{figure}[tb]
\vspace{-.7em}
\begin{minipage}{.5\linewidth}
  \centerline{\includegraphics[width=1.05\columnwidth]{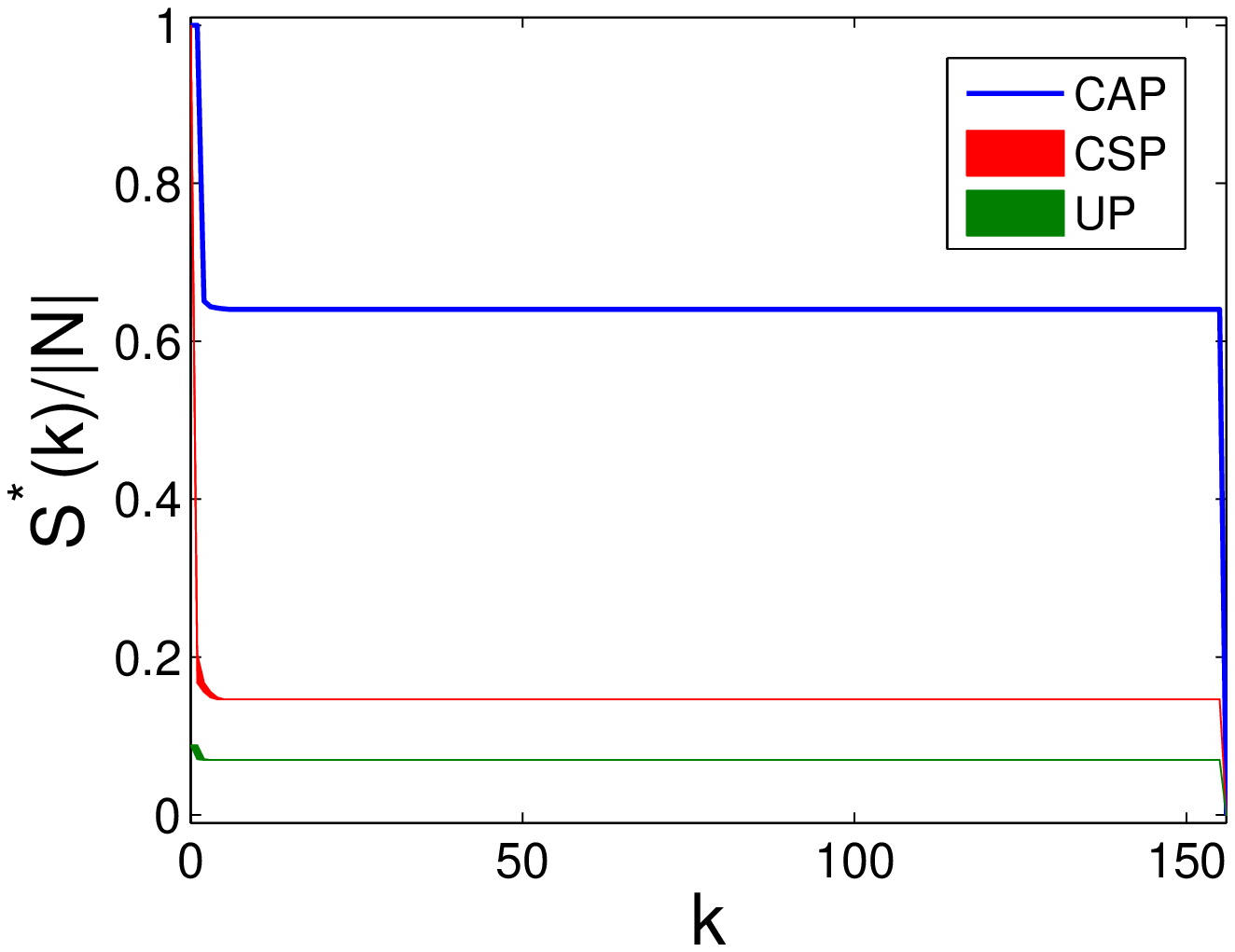}}
  \vspace{-.5em}\centerline{\scriptsize (a) $\mu=200$}
\end{minipage}\hfill
\begin{minipage}{.5\linewidth}
  \centerline{\includegraphics[width=1.05\columnwidth]{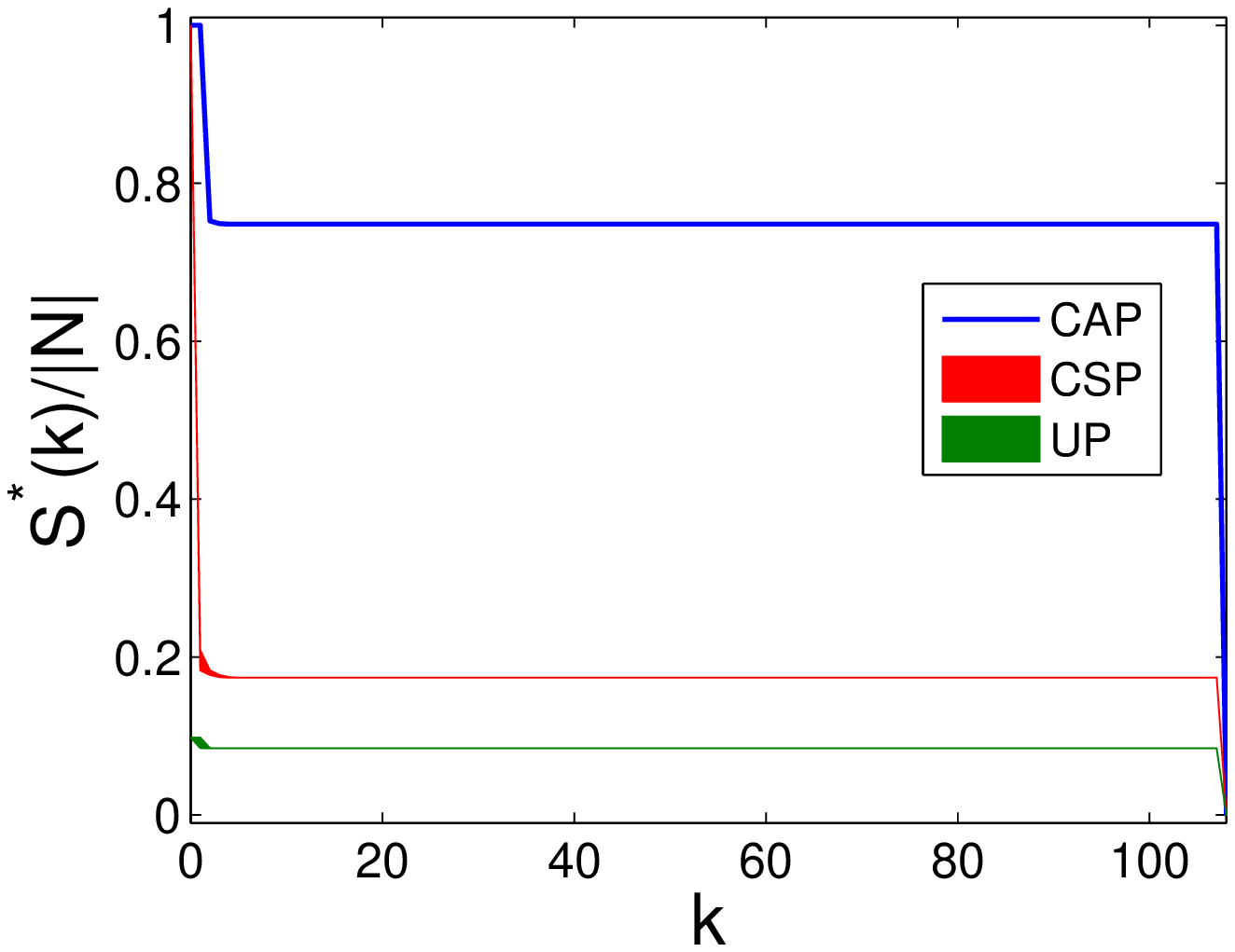}}
  \vspace{-.5em}\centerline{\scriptsize (b) $\mu=248$}
\end{minipage}
\vspace{-.5em}
\begin{minipage}{.5\linewidth}
  \centerline{\includegraphics[width=1.05\columnwidth]{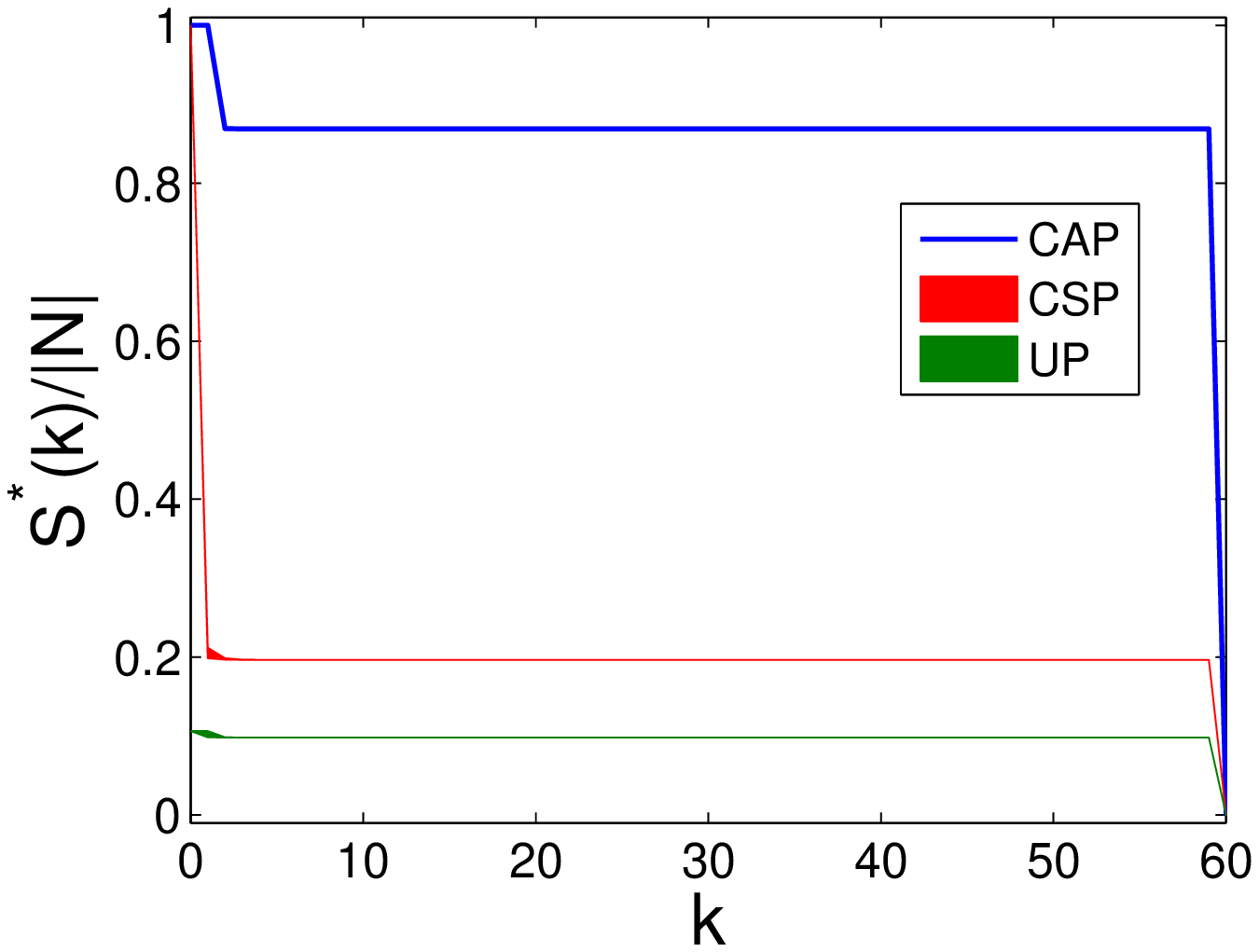}}
  \vspace{-.5em}\centerline{\scriptsize (c) $\mu=296$}
\end{minipage}\hfill
\begin{minipage}{.5\linewidth}
  \centerline{\includegraphics[width=1.05\columnwidth]{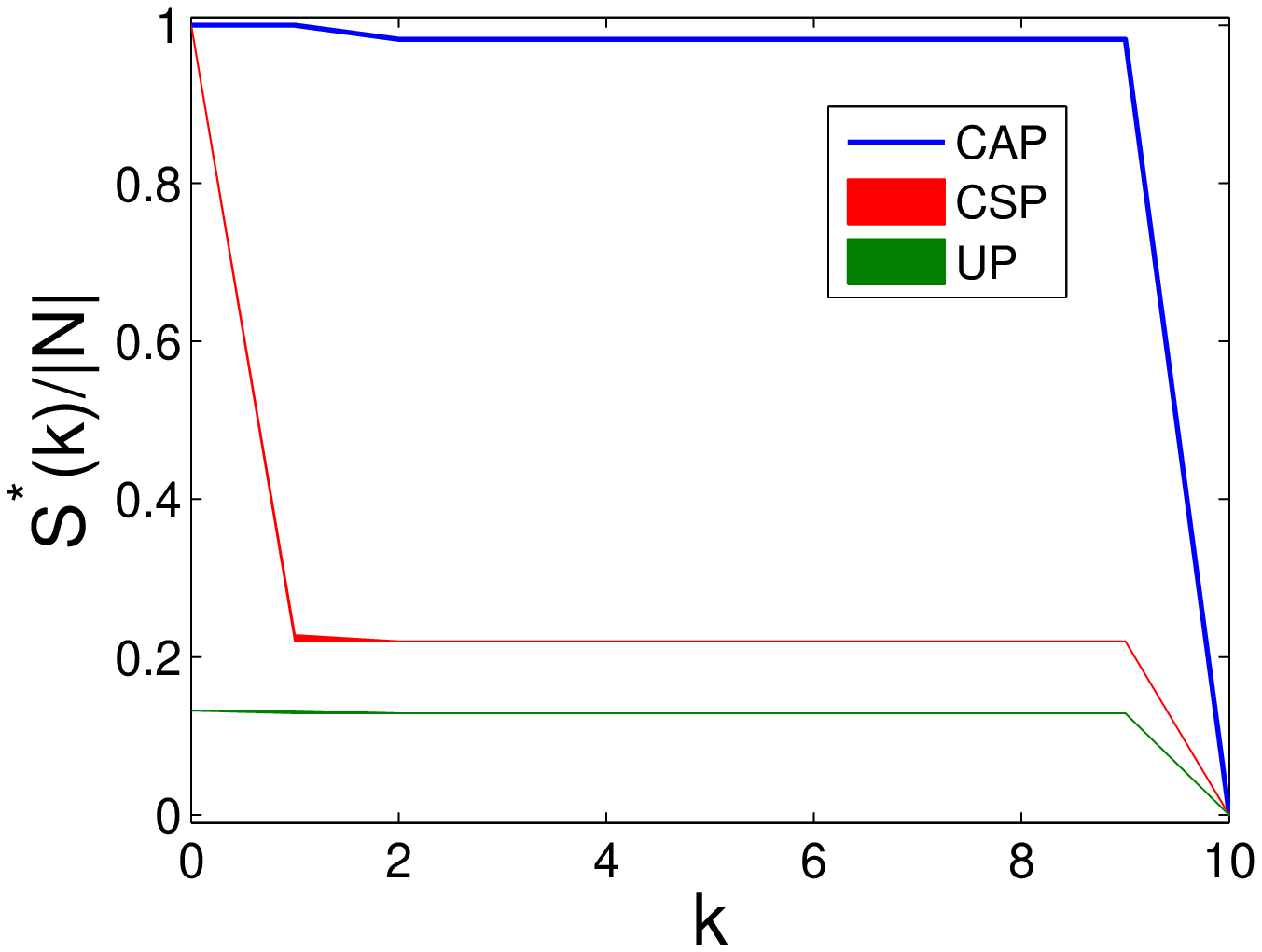}}
  \vspace{-.5em}\centerline{\scriptsize (d) $\mu=346$}
\end{minipage}
\vspace{-.0em}
\caption{Maximum $k$-identifiable set $S^*(k)$ under CAP, CSP, and UP for CAIDA AS26788 ($|V|=355$, $|L|=483$, $\mu=\{200,248,296,346\}$, $100$ Monte Carlo runs). } \label{fig:S_bounds_CAIDA}
\end{figure}

Fig.~\ref{fig:S_bounds_ER}--\ref{fig:S_bounds_CAIDA} show that under different graph models and monitor placements, similar observations/conclusions as Fig.~4--6 in \cite{Ma15TON} can be made.

\bibliographystyle{IEEEtran}
\bibliography{mybibSimplified,mybibSimplifiedA,mybibSimplifiedB,mybibSimplifiedC}
\end{document}